\documentclass[aps,twocolumn,prl,floatfix,superscriptaddress]{revtex4}
\usepackage{graphics,bm,amsmath,amssymb,natbib,url,epsfig,psfrag,color}
\usepackage{pdfpages}
\begin{document}

\title{Topological Vacuum Bubbles by Anyon Braiding}
\author{Cheolhee Han}
\affiliation{Department of Physics, Korea Advanced Institute of Science and Technology, Daejeon 305-701, Korea}
\author{Jinhong Park}
\affiliation{Department of Physics, Korea Advanced Institute of Science and Technology, Daejeon 305-701, Korea}
\author{Yuval Gefen}
\affiliation{Department of Condensed Matter Physics, Weizmann Institute of Science, Rehovot, 76100 Israel}
\author{H.-S. Sim}
\email[Corresponding author. ]{hssim@kaist.ac.kr}
\affiliation{Department of Physics, Korea Advanced Institute of Science and Technology, Daejeon 305-701, Korea}

\begin{abstract}

According to a basic rule of fermionic and bosonic many-body physics, known as the linked cluster theorem, physical observables are not affected by vacuum bubbles, which represent virtual particles created from vacuum and self-annihilating without interacting with real particles.
Here, we show that this conventional knowledge must be revised for anyons, quasiparticles that obey fractional exchange statistics intermediate between fermions and bosons. We find that a certain class of vacuum bubbles of Abelian anyons 
\emph{does} affect physical observables.
They represent virtually excited anyons which wind 
around real anyonic excitations.
These topological bubbles result in a temperature-dependent phase shift of Fabry-Perot interference patterns in the fractional quantum Hall regime accessible in current experiments, thus providing a 
tool for direct and unambiguous observation of elusive fractional statistics. 
\end{abstract}
\date{\today}
\maketitle



When two identical particles 
adiabatically exchange their positions $\textrm{\bf r}_\textrm{i=1,2}$, their final state $\psi$
(up to dynamical phase) 
is related to the initial one  through 
an exchange statistics phase $\theta^*$,
\begin{equation}
\psi (\textrm{\bf r}_2, \textrm{\bf r}_1) = e^{i \theta^*} \psi (\textrm{\bf r}_1, \textrm{\bf r}_2), \label{Exchange}
\end{equation}
with $\theta^*=0$ $(\pi)$ for bosons (fermions)~\cite{Fetter}.

Anyons~\cite{Leinaas,Arovas,Stern} are quasiparticles in two dimensions (2D),
not belonging to the two classes of elementary particles, bosons and fermions.
Abelian anyons appear in the fractional quantum Hall (FQH) system of filling factor $\nu = 1 / (2n+1)$, $n = 1,2,\cdots$. They carry a fraction $e^* = \nu e$ of the electron charge $e$,
and
obey fractional exchange statistics,
satisfying Eq.~\eqref{Exchange} with $\theta^* = \pm \pi \nu$. 
Two anyons gain a phase $\pm 2\pi\nu$ from a braiding whereby one winds around the other;
$\pm$ depends on the winding direction.
While fractional charges have been detected~\cite{Goldman,Picciotto,Saminadayar,Dolev}, 
experimental measurement of statistics phase $\pi \nu$
has been so far elusive. 
Existing theoretical proposals for the measurement involve quantities inaccessible in current experiments,
or suffer from unintended change of a proposed setup with external parameters~\cite{Chamon,Vishveshwara,Kim06,Safi,Campagnano12,Law,Feldman,Kane03,Rosenow,An}.


In many-body quantum theory~\cite{Fetter}, 
Feynman diagrams are used to compute the expectation value of observables.
This approach invokes vacuum bubble diagrams, which describe virtual particles excited from vacuum and self-annihilating without interacting with real particles.
According to the linked cluster theorem~\cite{Fetter}, each diagram possessing vacuum bubbles comes with, hence is exactly canceled by, a ``partner'' diagram of the same magnitude but of the opposite sign. Consequently, vacuum bubbles do not contribute to physical observables.


In the following, we demonstrate that this common wisdom has to be revised for anyons:
A certain class of vacuum bubbles of Abelian anyons 
\emph{does} affect observables.
These virtual particles, which we call ``topological vacuum bubbles'', wind around a real anyonic excitation, gaining the braiding phase $\pm 2 \pi \nu$. 
We propose a realistic setup for detecting
them and $\theta^* = \pi \nu$.

\vspace{.3cm}
\noindent {\bf Results}
\vspace{.1cm}

\paragraph*{\bf Topological Vacuum bubble.}
\begin{figure}
\centerline{\includegraphics[width=0.9\columnwidth]{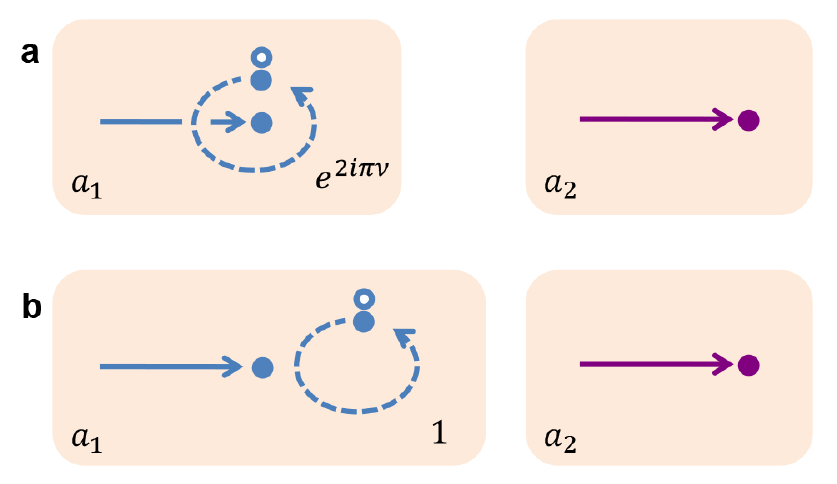}}
\caption{
{\bf Topological vacuum bubble.} Feynman diagrams for interference involving a real particle and a virtual particle-hole excitation from vacuum. Full (empty) circles represent particles (holes). Solid (dashed) lines denote propagations of real (virtual) particles.  
({\bf a}) Diagram for the interference $a_1 a_2^*$ of two processes: 
($a_1$, blue) 
A real particle propagates, a virtual particle-hole pair is excited, then the pair self-annihilates after the virtual particle winds around the real one. 
($a_2$, darkmagenta) A real particle propagates.
The entire virtual process constitutes a vacuum bubble. 
For anyons, the bubble gains a topological braiding phase $2\pi\nu$ from the winding.
({\bf b}) ``Partner'' diagram of ({\bf a}). Here a virtual particle, constituting another bubble, does not encircle a real one, hence, gains no braiding phase.
The diagrams ({\bf a}) and ({\bf b}) contribute to observables for anyons,
while they do not for bosons and fermions.
}
\label{fig1}
\end{figure}

We illustrate topological vacuum bubbles.
In Fig.~\ref{fig1}a, a Feynman diagram represents interference $a_1a_2^*$ between processes $a_1$ and $a_2$
of propagation of a real particle.
In $a_1$, 
a virtual particle-hole pair is excited then self-annihilates after the virtual particle winds around the real particle, forming a vacuum bubble, while it is not excited in $a_2$. 
%
The winding results in a braiding phase $2 \pi \nu$ and an Aharonov-Bohm phase $2 \pi \Phi / \Phi_0^*$ from the magnetic flux $\Phi$ enclosed by the winding path, contributing to the interference signal as $e^{i(2\pi\Phi/\Phi_0^*+2\pi\nu)}$; 
$\Phi_0^* = h/e^*$ is the anyon flux quantum~\cite{Kivelson,Chamon}.

The limiting cases of bosons ($\nu=0$) and fermions ($\nu=1$) imply that
this bubble diagram appears together
with, and is canceled by, a partner diagram in Fig.~\ref{fig1}b.
The partner diagram has a bubble not encircling the real particle, and involves only $2 \pi \Phi / \Phi_0^*$.
The two diagrams (and their complex conjugates) yield
\begin{align}
\textrm{Interference signal} &\propto  \textrm{Re} [e^{i (2 \pi \Phi/\Phi_0^* + 2 \pi \nu )} - e^{2 \pi i \Phi/\Phi_0^*}] \nonumber \\&= - \sin (\pi \nu) \sin (2 \pi \Phi/\Phi_0^* + \pi \nu). 
\label{Probability}
\end{align}
For bosons and fermions, the two diagrams fully cancel each other with $\sin(\pi\nu)=0$ in agreement with the linked cluster theorem, hence, the signal disappears. 
By contrast, for anyons, they cancel only partially, producing the nonvanishing interference
in an observable,
and are topological as the braiding phase is involved.


\begin{figure}
\includegraphics[width=0.93\columnwidth]{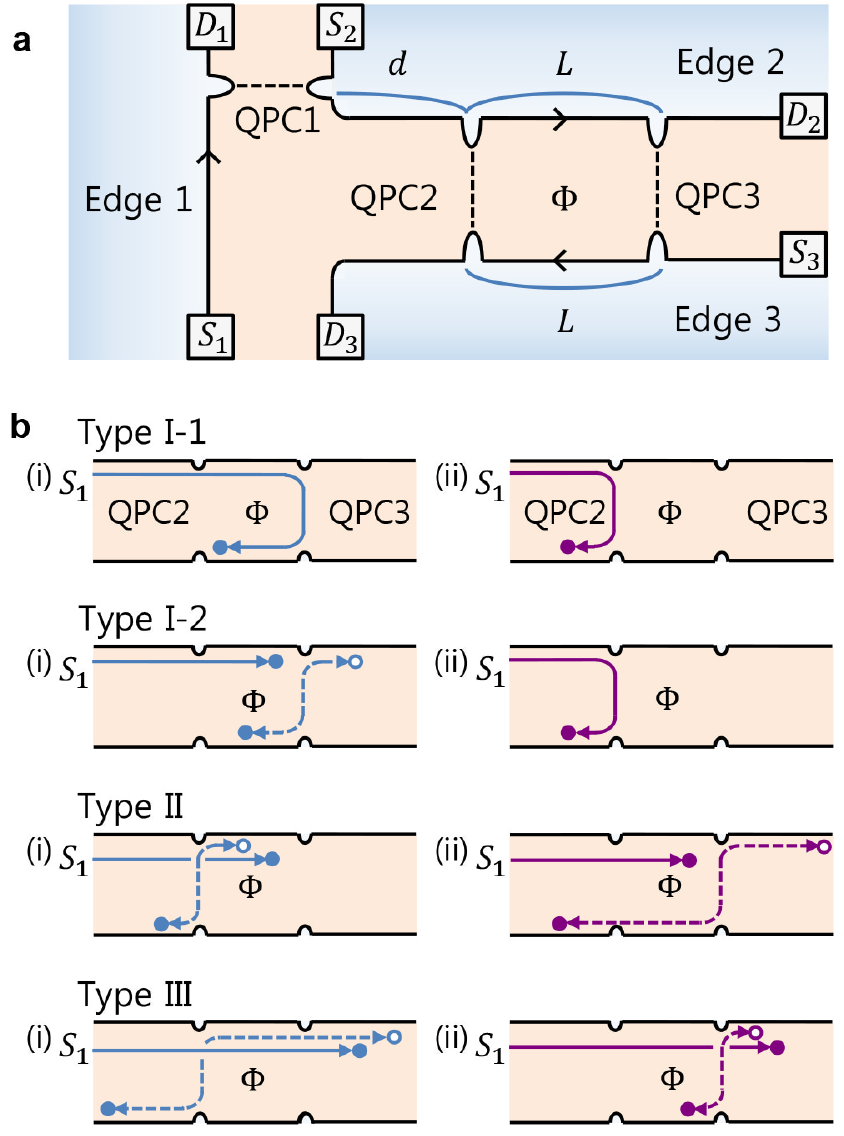}
\caption{
{\bf Interferometry for detecting topological vacuum bubbles.} ({\bf a}) 
In the setup, anyons move (see arrows) along fractional quantum Hall (FQH) edge channel $i=1,2,3$ that connects source S$_i$ and drain D$_i$, 
and jump (dashed) between the channels via tunneling at quantum point contacts (QPCs).
The loop defined by Edges $i=2,3$, QPC2, and QPC3 encloses magnetic flux $\Phi$, forming a Fabry-Perot interferometer. 
Distance between QPC2 and QPC3 (QPC1) is $L$ ($d$).
({\bf b}) Two interfering paths  (i) and (ii) of each main interference process at $e^* V \gg k_B T \gtrsim \hbar v_\textrm{p} / L$. 
Following Fig.~\ref{fig1}, 
filled (empty) circles represent particle-like (hole-like) anyons, 
and solid (dashed) lines denote propagation of an anyon injected from S$_1$ (anyon-pair excitation at QPCs). Type II and III processes involve a topological vacuum bubble. 
}
\label{fig2}
\end{figure}

\paragraph*{\bf Interferometer setup.}
In Fig.~\ref{fig2}a, we propose a minimal setup for observing topological vacuum bubbles.
It is a Fabry-Perot interferometer~\cite{Chamon,An,Willett,Camino,Ofek,McClure} in the $\nu = 1 / (2n + 1)$ FQH regime, coupled to an additional edge channel (Edge 1) via QPC1.  
At QPC$i$, there occurs tunneling of a single anyon (rather than anyon bunching), fulfilled~\cite{Kane92} with $\gamma_i \ll k_B T$; $\gamma_i$ is the tunneling strength and $T$ is temperature.
Gate voltage $V_\textrm{G}$ is applied, to change the interferometer loop enclosing Aharonov-Bohm flux $\Phi$.
The interference part $I_{\textrm{D}_3}^{\textrm{int}}$ of charge current at drain $D_3$
is measured
with bias voltage $V$ applied to source S$_1$;
the other S$_i$'s and D$_i$'s are grounded.  
Together with ``virtual'' (thermal) anyon excitations in the interferometer,
a voltage-biased ``real'' anyon, dilutely injected at QPC1 from Edge 1 to the interferometer,
forms topological vacuum bubbles, as shown below.  The bubbles contribute to $I_{\textrm{D}_3}^{\textrm{int}}$ at the leading order ($I_{\textrm{D}_3}^{\textrm{int}} \propto \gamma_1^2 \gamma_2 \gamma_3$) in QPC tunneling, since Edges 2 and 3 are unbiased.
Note that in the setups previously studied~\cite{Chamon,Vishveshwara,Kim06,Safi,Campagnano12,Law,Feldman,Kane03,Rosenow,An},
topological bubbles do not contribute to current at the leading order.

We consider the regime of $e^* V \gg k_B T \gtrsim \hbar v_\textrm{p} / L$, where the size $L_V \equiv \hbar v_\textrm{p} / (e^* V)$ of the dilutely injected anyons is much smaller than interferometer size $L$ and the injection of hole-like anyons at QPC1 is ignored; $v_\textrm{p}$ is anyon velocity along the edges and $e^*V$ should be much smaller than the FQH energy gap.
Because of the dilute injection and $L_V \ll L$, anyon braiding is well defined in the interferometer. As shown below, the dependence of $I_{\textrm{D}_3}^{\textrm{int}}$  on $\Phi$ or on $V_\textrm{G}$ provides a clear signature of the topological bubbles, consequently, $\theta^* = \pi \nu$ in both of the pure Aharonov-Bohm regime (where Coulomb interaction of the edge channels with bulk anyons localized inside the interferometer loop is negligible) and the Coulomb dominated regime (where the interaction is strong)~\cite{Ofek,Halperin}. Below, we first ignore bulk anyons.


\paragraph*{ \bf Interference current.}
Employing the chiral Luttinger liquid theory~\cite{Delft,Wen} for FQH edges and Keldysh Green's functions~\cite{Safi,Kim06}, we compute $I_{\textrm{D}_3}^{\textrm{int}}$ ($\propto \gamma_1^2 \gamma_2 \gamma_3$) at the leading order in $\gamma$'s.
There are four types of the processes mainly contributing to $I_{\textrm{D}_3}^{\textrm{int}} \simeq I_{\textrm{D}_3}^{\textrm{I-1}} + I_{\textrm{D}_3}^{\textrm{I-2}} + I_{\textrm{D}_3}^{\textrm{II}} + I_{\textrm{D}_3}^{\textrm{III}}$; see Fig.~\ref{fig2}b. 
For $e^* V \gg k_B T \gg \hbar v_\textrm{p} / L$, we obtain the analytical
expression of $I_{\textrm{D}_3}^{\textrm{int}}$: The interference current contributed by Type I-1 processes is
$I_{\textrm{D}_3}^{\textrm{I-1}} \propto  V^{\nu-2} T^{1-\nu} g(V,T) \cos(2\pi\Phi/\Phi_0^*+e^* V L/\hbar v_\textrm{p})$,  that by Type I-2 is
$I_{\textrm{D}_3}^{\textrm{I-2}} \propto V^{-1}  g(V,T) (\sin^2\pi\nu) \cos(2\pi\Phi/\Phi_0^*)$, and those by Type II and III are
\begin{align}
I_{\textrm{D}_3}^{\textrm{II}}\,  \simeq &\,  g(V,T)
\frac{2L + C (\nu) L_T}{\hbar v_\textrm{p}} (\sin^2 \pi\nu) \cos(2\pi\Phi/\Phi_0^{*}+\pi\nu), \label{Analytic_I} \\ 
I_{\textrm{D}_3}^{\textrm{III}}\, \simeq &\, g(V,T)
\frac{C (\nu) L_T}{\hbar v_\textrm{p}} (\sin^2\pi\nu) \cos(2\pi\Phi/\Phi_0^{*}-\pi\nu)\label{Analytic_II}, 
\end{align} 
where $g(V,T) \propto \gamma_1^2 \gamma_2 \gamma_3  (V T)^{2\nu-1} e^{-2L/L_T}$, $e^{-2L/L_T}$ is a thermal suppression factor, thermal length $L_T \equiv \hbar v_\textrm{p} / (\pi \nu k_B T)$, and
$C(\nu = 1/3) \simeq 0.43$; see the Supplementary Note 3.

Type I-1 processes describe interference between two paths of an anyon moving from S$_1$ to D$_3$ via (i) QPC2 and (ii) QPC3, respectively. They were previously studied~\cite{Chamon}. 

In Type I-2, an anyon injected from S$_1$ interferes with a particle-like anyon excited at a QPC, or annihilates a hole-like anyon; particle-like and hole-like anyons are pairwise excited thermally at QPCs.
For example, consider the two following interfering histories: (i)
An anyon is injected from S$_1$ to D$_2$, an anyon pair is excited at QPC3, and then the particle-like (hole-like) anyon of the pair moves to D$_3$ (D$_2$). 
(ii) An anyon is injected from S$_1$  to D$_3$ via QPC2 without any excitations. 
The hole-like anyon annihilates the injected anyon on Edge~2 in history~(i), and the particle-like anyon of (i) interferes with the injected anyon of (ii) on Edge~3.
The sum of such interference processes yields $I_{\textrm{D}_3}^{\textrm{I-2}} \propto \sin^2\pi\nu \cos (2 \pi \Phi/ \Phi_0^*)$. 
The $\sin^2 \pi \nu$ factor appears
because relative locations of anyons on Edge 2 or 3 differ between the processes, leading to an
exchange phase $\pm \pi \nu$, and because a process with an excitation (of a particle-like anyon moving to D$_2$ and a hole-like one to D$_3$)
yields charge current in the opposite direction to another with its particle-hole conjugated excitation (of a particle to D$_3$ and a hole to D$_2$).


In Types II and III, a ``real'' anyon injected from S$_1$ moves to D$_2$, and a ``virtual'' anyon pair excited at QPC2 interferes with another at QPC3. 
The interference path effectively encloses the real anyon, forming a topological vacuum bubble; cf. Fig.~\ref{fig1}.
In Type II, when the real anyon is located on Edge~2 between QPC2 and QPC3, a virtual pair is excited at QPC2 in history (i), and at QPC3 in history (ii). Then the hole-like (particle-like) anyon of each pair moves, for example, to D$_2$ (D$_3$). 
The interference of the two histories corresponds to 
the winding of a virtual anyon around the real one and $\Phi$, forming
a topological bubble with interference phase $\pm (2 \pi \Phi/\Phi_0^* + 2 \pi \nu)$; $\pm$ depends on whether the hole-like anyon moves to D$_2$ or D$_3$.
In the interference, the winding of a virtual anyon around the real one effectively occurs through the exchanges of the positions of the anyons in each of Edges 2 and 3, as relative locations of anyons on Edges 2 and 3 differ between (i) and (ii); see Supplementary Figure 2 and Supplementary Note 6.
This interference is accompanied by a partner process. 
The latter has a bubble that winds around $\Phi$ (gaining phase $\pm 2 \pi \Phi/\Phi_0^*$), but not around a real anyon.
The two partner processes  partially cancel each other, yielding
$I_{\textrm{D}_3}^{\textrm{II}} \propto \sin \pi \nu \cos (2 \pi \Phi/\Phi_0^* + \pi \nu)$;
the remaining $\sin \pi \nu$ factor in Eqs.~\eqref{Analytic_I} and~\eqref{Analytic_II} has a similar origin to
the $\sin^2 \pi \nu$ factor of $I_{\textrm{D}_3}^{\textrm{I-2}}$.

In Type III, the two interfering histories are: (i) A virtual pair is excited at QPC2 before a real anyon injected from S$_1$ arrives at QPC2, and (ii) another pair is excited at QPC3 after the real one arrives at QPC3. 
The ensuing chronological sequence on Edge~2 is opposite to Type II: An anyon excited at QPC2 arrives at QPC3; the real one arrives at QPC3; a pair is excited at QPC3.
The resulting topological bubble effectively winds around the real anyon in the direction opposite to its winding around $\Phi$,
yielding a phase $\pm (2 \pi \Phi/\Phi_0^* - 2 \pi \nu)$.
Partial cancellation of the bubble and its partner leads to $I_{\textrm{D}_3}^{\textrm{III}} \propto  \sin \pi \nu \cos (2 \pi \Phi/\Phi_0^* - \pi \nu)$.
The factor $2L + C L_T$ of $I_{\textrm{D}_3}^{\textrm{II}}$
($C L_T$ in $I_{\textrm{D}_3}^{\textrm{III}}$) in Eq.~\eqref{Analytic_I} (Eq.~\eqref{Analytic_II}) comes from the time window compatible with the chronological sequence on Edge~2.



Type II and III processes of topological bubbles 
do not affect any observables at $\nu=1$ (fermions),
due to
full cancellation between partner bubbles (the linked cluster theorem).
They are distinct from I-2. 
I-2 processes produce, e.g., nonvanishing current noise $\langle (I_{\textrm{D}_3}^{\textrm{int}} - \langle I_{\textrm{D}_3}^{\textrm{int}} \rangle)^2\rangle$ at $\nu=1$, as the particle-hole conjugated excitations (mentioned before) equally contribute to the noise
(while the contributions of the conjugations to $I_{\textrm{D}_3}^{\textrm{I-2}}$
cancel each other, leading to $I_{\textrm{D}_3}^{\textrm{I-2}} = 0$).

In the more general regime of $e^* V \gg k_B T \gtrsim \hbar v_\textrm{p} / L$, we employ the parametrization
\begin{eqnarray}
I_{\textrm{D}_3}^{\textrm{int}} \propto \cos(2\pi\Phi/\Phi_0^*+\theta). \label{Interference}
\end{eqnarray}
The phase $\theta$ is determined by competition between the various contributions to $I_{\textrm{D}_3}^{\textrm{int}}$ and contains information about statistics phase $\pi \nu$. 
At $e^* V \gg k_B T \gtrsim \hbar v_\textrm{p} / L$, $I_{\textrm{D}_3}^{\textrm{II}} + I_{\textrm{D}_3}^{\textrm{III}}$ is much larger than $I_{\textrm{D}_3}^{\textrm{I-1}} + I_{\textrm{D}_3}^{\textrm{I-2}}$ and dominates $I_{\textrm{D}_3}^{\textrm{int}}$, because
the interfering anyon of Types I-1 and I-2 is voltage biased and has width $L_V \propto V^{-1}$ much narrower than the thermal anyon excitations (whose width $L_T \propto T^{-1}$) of II and III, showing much weaker interference.
From $I_{\textrm{D}_3}^{\textrm{II}}$ and $I_{\textrm{D}_3}^{\textrm{III}}$,
we find
\begin{align}
\theta & \simeq  \arctan [ \frac{L}{L + C(\nu) L_T} \tan (\pi \nu + (1-2 \nu) \arctan \frac{2L_V}{L_T} ) ] \label{phase_shift} \\
& \to \arctan [ \frac{L}{L + C(\nu) L_T} \tan \pi \nu ] \quad \quad \quad \textrm{as} \,\,T/V \to 0  \nonumber \\
& \to  \pi \nu \quad \quad \quad \textrm{as} \,\, L_T / L \to 0 \,\, \textrm{ and } \,\, T/V \to 0. \nonumber
\end{align}
The $\arctan 2 L_V / L_T$ term represents an error in the braiding phase $2 \pi \nu$ of the topological bubbles. 
It occurs when
the size $L_V$ of a real anyon is not sufficiently smaller than 
the winding radius of a virtual anyon around the real one.
It is negligible at $e^* V \gg k_B T$, since
the radius is effectively large when $L_V \ll L_T$; those corresponding to the error are ignored in Eqs.~\eqref{Analytic_I} and~\eqref{Analytic_II}, and found in Supplementary Note 3. 
For $e^* V \gg k_B T \gg \hbar v_\textrm{p} / L$, $I_{\textrm{D}_3}^{\textrm{II}}$ dominates $I_{\textrm{D}_3}^{\textrm{int}}$ and $\theta \to \pi \nu$.
Remarkably, $\theta$ depends on $T$, contrary to common practice in electron interferometry~\cite{Ji}. 

\begin{figure}
\includegraphics[width=\columnwidth]{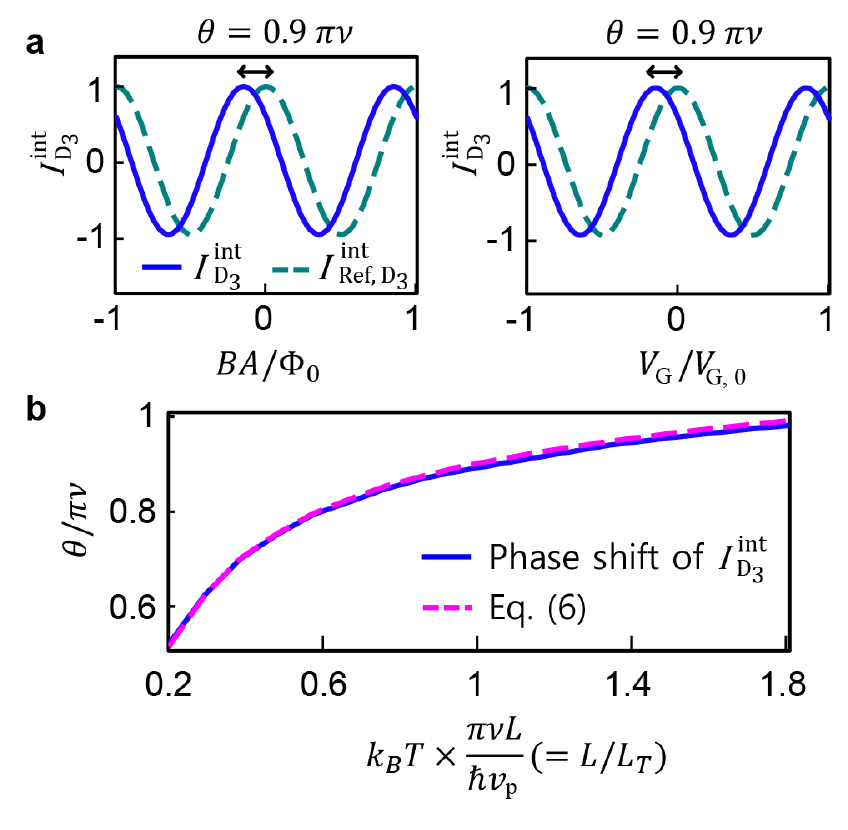}
\caption{
{\bf Detection of anyon phase $\pi\nu$ from interference phase shift $\theta$.} ({\bf a}) Dependence of $I_{\textrm{D}_3}^{\textrm{int}}$ (blue, normalized) and $I^{\textrm{int}}_{\textrm{Ref, D}_3}$ (darkcyan, normalized) on $\Phi$ in the pure Aharonov-Bohm regime (left panel) and their dependence on $V_{\textrm{G}}$ in the Coulomb dominated regime (right).
We choose $\nu=1/3$, $T=30 \,m\textrm{K}$, $e^* V = 45 \, \mu e \textrm{V}$, $L=3\,\mu\textrm{m}$, and $v_\textrm{p} = 10^4$ m~s$^{-1}$ ($ L / L_T = 1.2$ and $L/ L_V = 20$); see Supplementary Note 5 for the Coulomb-interaction parameter of the regimes.
For these parameters, the phase shift $\theta$ between $I_{\textrm{D}_3}^{\textrm{int}}$ and $I^{\textrm{int}}_{\textrm{Ref, D}_3}$ is $0.9 \pi \nu$.
({\bf b}) Dependence of $\theta$ on $T$. The same parameters (except $T$) with ({\bf a}) are chosen.
$\theta \to \pi\nu$ as $T$ increases (yet $e^* V \gg k_B T \gg \hbar v_\textrm{p} / L$).
In both the pure Aharonov-Bohm regime and the Coulomb dominated regime,
the same numerical result (blue curve) of $\theta(T)$, which
agrees with Eq.~\eqref{phase_shift} (darkmagenta), is obtained.
}
\label{fig3}
\end{figure}

\paragraph*{\bf Coulomb dominated regime.}
In experimental situations of a Fabry-Perot interferometer in the FQH regime, it is expected that there exist bulk anyons localized inside the interferometer loop.
There are two regimes of Fabry-Perot interferometers, the pure Aharonov-Bohm regime and the
Coulomb dominated regime.
In the former regime, Coulomb interaction between the bulk anyons and the edge of the interferometer is negligible, while it is crucial in the latter~\cite{Halperin}.
The Fabry-Perot interferometers of recent experiments~\cite{An,Willett,Camino,Ofek,McClure} in the FQH regime are in the Coulomb
dominated regime.
We below compute the interference current $I_{\textrm{D}_3}$ in the presence of the Coulomb interaction, and show that Eq.~\eqref{phase_shift} is applicable to both of the pure Aharonov-Bohm regime and the Coulomb
dominated regime. 

For $e^* V \gg k_B T \gtrsim \hbar v_\textrm{p} / L$, we numerically compute $I_{\textrm{D}_3}^{\textrm{int}}$ ($\propto \gamma_1^2 \gamma_2 \gamma_3$) in Fig.~\ref{fig3},
combining our theory with the capacitive interaction model~\cite{Halperin} that successfully describes thermally fluctuating bulk anyons and the interaction; see Supplementary Note 5.
We find the gate-voltage dependence of
$I_{\textrm{D}_3}^{\textrm{int}} \propto \cos(2 \pi V_\textrm{G}/V_{\textrm{G},0}+\theta)$ with periodicity $V_{\textrm{G},0}$
in the Coulomb dominated regime, and $I_{\textrm{D}_3}^{\textrm{int}} \propto \cos(2 \pi \Phi/\Phi_{0}+\theta)$  in the pure Aharonov-Bohm limit; here the periodicity of the $\Phi$ dependence is $\Phi_0 \equiv h/e$ rather the period $\Phi_0^*$ of Eq.~\eqref{Interference}, because of the fluctuation of the number of bulk anyons~\cite{Kivelson,Chamon}.
In both the regimes, the interference processes discussed before (Fig.~\ref{fig2}) appear in the same manner, hence, $\theta$ satisfies the analytic expression in Eq.~\eqref{phase_shift}; cf. Fig.~\ref{fig3}b.


\paragraph*{\bf How to measure the phase $\theta$.}
Experimental measurements of $\theta$ can be affected by possible side effects, including the external-parameter (magnetic fields, gate voltages, and bias voltages) dependence of the size, shape, QPC tunneling, and bulk anyon excitations.
Below, we propose how to detect $\theta$ with avoiding the side effects, utilizing the setup in Fig.~\ref{fig2}a.

The phase $\theta$ is experimentally measurable, 
by comparing $I_{\textrm{D}_3}^{\textrm{int}}$ with a reference current $I^{\textrm{int}}_{\textrm{Ref, D}_3}$.
$I^{\textrm{int}}_{\textrm{Ref, D}_3}$ is measured at D$_3$ in the same setup under the same external parameters (temperature, gate voltages, magnetic field, etc) with $I^{\textrm{int}}_{\textrm{D}_3}$, but with applying infinitesimal bias voltage $V_\textrm{ref} / 2$ to S$_2$ and $-V_\textrm{ref}/2$ to S$_3$ and keeping S$_1$ and all D$_i$'s grounded~\cite{Chamon}; cf. Supplementary Note 4.
In any regimes $I_{\textrm{D}_3}^{\textrm{int}}$ shows the same interference pattern with $I^{\textrm{int}}_{\textrm{Ref, D}_3}$, but is phase-shifted from $I^{\textrm{int}}_{\textrm{Ref, D}_3}$ by $\theta$;
$I^{\textrm{int}}_{\textrm{Ref, D}_3} \propto \cos 2 \pi V_\textrm{G}/ V_{\textrm{G},0}$ ($I^{\textrm{int}}_{\textrm{Ref, D}_3} \propto \cos 2 \pi \Phi / \Phi_0$) in the Coulomb dominated (pure Aharonov-Bohm) limit. 
Importantly, the side effects 
modify $I_{\textrm{D}_3}^{\textrm{int}}$ and $I^{\textrm{int}}_{\textrm{Ref, D}_3}$ in the same manner,
hence, the phase shift between the patterns remains as $\theta$.

The fractional statistics phase is directly and unambiguously identifiable in experiments, by observing $\theta \to \pi \nu$ at $e^* V \gg k_B T \gg \hbar v_\textrm{p} / L$ with excluding the side effects as above. Or, one applies the fit function of $\arctan [ A_1 / (1 + A_2/T)]$ with fit parameters $A_1$ and $A_2$ to measured data of $\theta(T)$, and extracts $\arctan A_1 = \pi \nu$ from the fit; cf. the second line of Eq.~\eqref{phase_shift}.
Observation of $\theta = \pi \nu$ or $\theta(T)$ will suggest a strong evidence of anyon braiding and topological bubbles.

The parameters in Fig.~\ref{fig3} are experimentally accessible~\cite{An,Camino,Ofek,McClure,Willett}. 
For the QPCs, there are constraints (i) that the number of voltage-biased anyons injected through QPC1 is at most one in the interferometer loop at any instance (to ensure that the braiding phase of a topological vacuum bubble is $2 \pi \nu$), (ii) that the anyon tunneling probabilities at QPC2 and QPC3 are sufficiently small (to ensure that the double winding of an anyon along the interferometer loop is negligible), and (iii) that anyon tunneling (rather than electron tunneling) occurs at the QPCs.
The constraint (i) is satisfied when the anyon tunneling probability at QPC1 is less than $\hbar v_\textrm{p} / (2 L e^* V)$, which is about 0.05 under the parameters.
To achieve the constraints (ii) and (iii), each tunneling probability of QPC2 and QPC3 is typically set to be 0.4 in experiments~\cite{Ofek,Griffiths}; then the amplitude of the double winding is smaller than that of the single winding by the factor $0.4 \exp (-2L / L_T)$ which is about 0.04 at 30 mK.
With the constraints we estimate the amplitude of $I_{\textrm{D}_3}^{\textrm{int}} \lesssim (\nu e^2 V / h) (\hbar v_\textrm{p} / (2 L e^* V)) 0.4 \exp(-2L / L_T)$, which is 1.5 pA at 30 mK and 0.6 pA at 40 mK under the parameters. Note that $\theta = 0.9 \pi \nu$ is reached at 30 mK while $\theta = 0.95 \pi \nu$ at 40 mK under the parameters.
The estimation is within a measurable range in experiments, where current $\gtrsim 0.5$ pA is well detectable~\cite{Comforti}.

We remark that the above strategy of detecting the fractional statistics phase is equally applicable to the more general quantum Hall regime~\cite{Ofek} of filling factor $\nu' = \nu + \nu_0$, in which the edge channels from the integer filling $\nu_0$ are fully transmitted through the QPCs while the channel from the fractional filling $\nu$ forms the interferometry in Fig.~\ref{fig2}. For this case we compute $I_{\textrm{D}_3}^{\textrm{int}}$ and $I^{\textrm{int}}_{\textrm{Ref, D}_3}$,
and find that in both of the pure Aharonov-Bohm regime and the Coulomb dominated regime the interference-pattern phase shift between them is identical to the phase $\theta$ of the $\nu_0 = 0$ case discussed in Eq.~\eqref{phase_shift} and Fig.~\ref{fig3}; cf. Supplementary Note 5.

Certain anyonic vacuum bubbles involve topological braiding and affect physical observables
surprisingly, contrary to vacuum bubbles of bosons and fermions.
They can be detected with current experiment tools, which will provide an unambiguous evidence of anyonic fractional statistics.
We expect that they are relevant also for 
other filling fractions $\nu = p / (2np + 1)$, non-Abelian anyons~\cite{An,Willett}, and 
topological quantum computation setups~\cite{Nayak}.

\vspace{.3cm}
\noindent {\bf Methods}
\vspace{.1cm}

\paragraph*{\bf Hamiltonian for the interferometer.}
We present the Hamiltonian for the setup. We recall the chiral Luttinger liquid theory for fractional quantum Hall edges.

The Hamiltonian $H = \sum_i H_{\textrm{edge}, i} + H_\textrm{tun}$ for the interferometer in Fig.~\ref{fig2}a
consists of $H_{\textrm{edge}, i}$ for edge channel $i$ and $H_\textrm{tun}$ for anyon tunneling at QPCs. Edge channel 1 is biased by $V$, and its Hamiltonian, employing the bosonization~\cite{Delft} 
for chiral Luttinger liquids, is given by
\begin{equation}
H_{\textrm{edge},1} =\frac{\hbar v_\textrm{p}}{4\pi \nu} \int_{-\infty}^{\infty} dx :(\partial_x\phi_1(x))^2:+e^* V \hat{N}_1.
\end{equation}
For the other unbiased channels, $H_{\textrm{edge},i=2,3} = (\hbar v_\textrm{p} / 4\pi \nu) \int_{-\infty}^{\infty} dx :(\partial_x\phi_i(x))^2:$.
Here, $e^*>0$, $\hat{N}_i$ is the anyon number operator of channel $i$, and $\phi_i (x)$ is the bosonic field of channel $i$ at position $x$, which describes the plasmonic excitation of anyons. 
The tunneling Hamiltonian is $H_\textrm{tun} = T^{\rightarrow}_1+T^{\downarrow}_2+T^{\downarrow}_3+\textrm{h.c.}$. $T^{\rightarrow}_1$ is the operator from Edge channel 1 to 2 at QPC 1, $T^{\downarrow}_2$ from Edge 2 to 3 at QPC2, and $T^{\downarrow}_3$ from Edge 2 to 3 at QPC3. These are written as
\begin{align}
 T_1^{\rightarrow}(t)&=\gamma_1 \Psi_2^{\dagger}(0,t)\Psi_1(0,t), \\
 T_2^{\downarrow}(t)&=\gamma_2 e^{-i\pi\Phi/\Phi_0^*} \Psi_3^{\dagger}(L,t)\Psi_2(d,t),\\
T_3^{\downarrow}(t)&=\gamma_3 e^{i\pi\Phi/\Phi_0^*} \Psi_3^{\dagger}(0,t)\Psi_2(d+L,t),
\end{align}
where $\Psi_i^{\dagger}(x,t) = F_i^\dagger (t) e^{- i \phi_i (x,t)} / \sqrt{2 \pi a}$  creates an (particle-like) anyon at position $x$ and time $t$ on Edge $i$, $a$ is the short-length cutoff, $\gamma_i$ is the tunneling strength at QPC $i$ (chosen as real), and the Aharonov-Bohm flux $\Phi$ enclosed by Edges $i=2,3$, QPC 2, and QPC3 is attached to $T_2^{\downarrow}$ and $T_3^{\downarrow}$ under certain gauge transformation; the dynamical phase common to the all edge channels is absorbed to $2\pi \Phi / \Phi_0^*$.
The Klein factor $F_i^\dagger$ increases the number of anyons on Edge~$i$ by one, and satisfies
$F_i^\dagger F_i =F_i F_i^\dagger =1$, $[N_i,F^\dagger_j]=\delta_{ij} F^\dagger_i$, 
$F_1(t)=F_1(0) e^{-ie^* V t / \hbar}$, and $[\phi_i, F_j] = 0$. 

\begin{figure}
\includegraphics[width=\columnwidth]{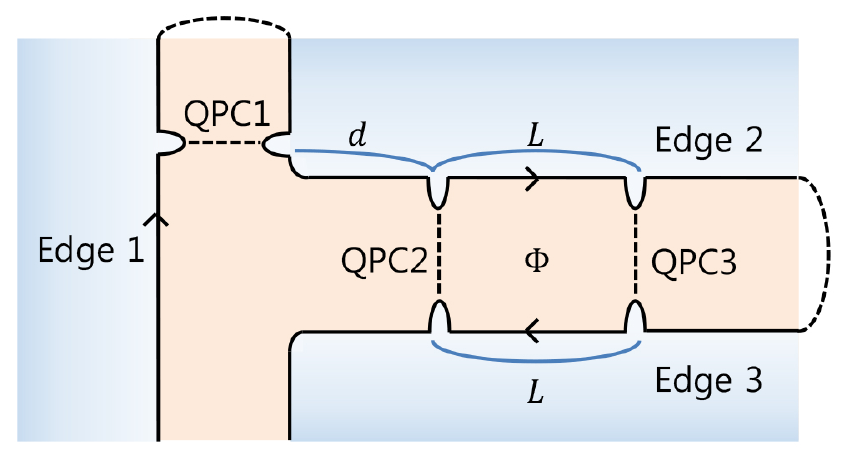}
\caption{
{\bf Extended edge channel scheme.} It is obtained by connecting the edge channels of the setup in Fig.~2a. The connection is represented by dashed arcs, while anyon propagation direction and anyon tunneling at quantum point contacts (QPCs) by arrows and dashed lines, respectively.
}
\label{geometry_connected}
\end{figure}

The exchange rule in Eq.~\eqref{Exchange} is described by $\phi_i$ and $F_i$. On Edge $i$, it is satisfied by
\begin{equation}
[\phi_i (x_1 ),\phi_j (x_2 )]= i\pi\nu \delta_{ij} \textrm{sgn}(x_1 -x_2 ).
\label{bosonic_commutator} \end{equation}
The exchange rule between anyons on different edges is achieved with the commutators of $F_i$, 
\begin{equation}
F_i F_j =F_j F_i e^{-i \pi\nu \textrm{sgn}(i-j)}, \quad \quad F_i^\dagger F_j =F_j F_i^\dagger e^{i \pi\nu\textrm{sgn}(i-j) }.
\label{Klein_commute} \end{equation}
A conventional way~\cite{Guyon}  
for obtaining the commutators
is to think of an extended edge connecting the different channel segments with no twist (cf. Fig.~\ref{geometry_connected}). The connection should preserve the chiral propagation direction of the channels. The exchange rule $\Psi(x_1)\Psi(x_2)=\Psi(x_2)\Psi(x_1)e^{-i\pi\nu\textrm{sgn}(x_1-x_2)}$ of anyons of the extended edge agrees with Eqs.~\eqref{bosonic_commutator} and \eqref{Klein_commute}.



We consider the regime of weak tunneling of anyons, and treat $H_\textrm{tun}$ as a perturbation on $\sum_i H_{\textrm{edge}, i}$. Perturbation theory is applicable~\cite{Kane92}
 in the renormalization group sense, when 
$e^* V$ and $k_B T$ are higher than $C \gamma^{1 / (1-\nu)}$, $C$ being a non-universal constant. 


The current $I_{\textrm{D}_3}$ is expressed as $I_{D_3} = i e^* [N_3, H] = i e^* ( T_2^\downarrow - T_2^\uparrow  + T_3^\downarrow - T_3^\uparrow )$.
$I_{\textrm{D}_3}$ is decomposed,
$I_{\textrm{D}_3 } = I_{\textrm{D}_3}^{\textrm{dir}} + I_{\textrm{D}_3}^{\textrm{int}}$, into direct current $I_{\textrm{D}_3}^{\textrm{dir}} \propto \gamma_1^2 \gamma_2^2, \, \gamma_1^2 \gamma_3^2$, and interference current $I_{\textrm{D}_3}^{\textrm{int}} \propto \gamma_1^2 \gamma_2 \gamma_3$ depending on $\Phi$ (the leading-order contribution).  Equations~\eqref{Analytic_I} and~\eqref{Analytic_II} are obtained by employing Keldysh Green's function technique with semiclassical approximation; see Supplementary Note 1, 2 and 3.

\paragraph*{\bf Coulomb interaction.} 
In presence of Coulomb interaction between bulk anyons and edge channels, 
we compute $I^{\textrm{int}}_{\textrm{D}_3}$, combining our chiral Luttinger liquid theory with the capacitive interaction model~\cite{Halperin} that successfully describes the Coulomb dominated regime. 
The interferometer Hamiltonian $H = \sum_i H_{\textrm{edge}, i} + H_\textrm{tun}$ is modified by the Coulomb interaction as
\begin{align}
H\to & H+U_\textrm{bulk} Q_{\textrm{bulk}}^2\nonumber\\&+U_{\textrm{int}}Q_{\textrm{bulk}} \int_{0}^{L} dx(:\partial_x \phi_2(x+d):+:\partial_x \phi_3(x): )\nonumber\\
=&\sum_{i=1,2,3}\frac{\hbar v_\textrm{p}}{4\pi\nu}\int_{-\infty}^{\infty} dx: (\partial_x \bar{\phi}_i(x))^2:+ H_{\textrm{bulk}} + H_\textrm{tun}.\label{NEWH}
\end{align}
Here, $Q_{\textrm{bulk}}=\nu B A_{\textrm{area}}/\Phi_0 +\nu N_L -\bar{q}$ is the number of the net charges localized within the interferometer bulk (inside the interference loop), $A_\textrm{area}$ is the area of the interferometer, $N_L$ is the net number of quasiparticles minus quasiholes, and $\bar{q}$ is the number of positive background charges induced by the gate voltage applied to the interferometer. $U_\textrm{int}$ is the strength of Coulomb interaction between the charges of the interferometer edge and the charges localized in the interferometer bulk, and $U_\textrm{bulk}$ is the strength of interaction between the bulk charges. 
In the second equality of Eq.~\eqref{NEWH}, we introduce a boson field $\bar{\phi}_i$ for each Edge $i$, $\bar{\phi}_i(x)=\phi_i(x)+\frac{2\pi\nu}{\hbar v_\textrm{p}} U_\textrm{int} Q_{\textrm{bulk}} \int_{-\infty}^{x} K_i(x')dx'$,
where $K_2(x)= 1$ for $d<x<d+L$, $K_3(x) = 1$ for $0<x<L$, and $K_i(x) = 0$ otherwise.
The second term of $\bar{\phi}$ describes the charges
$-\frac{2\pi\nu}{\hbar v_\textrm{p}} U_\textrm{int} Q_{\textrm{bulk}}$ induced per unit length by the interaction.
In Eq.~\eqref{NEWH}, the Hamiltonian is quadratic in $\bar{\phi}_i$ and has $H_{\textrm{bulk}}= (U_\textrm{bulk}-\frac{2\pi\nu L}{\hbar v_\textrm{p}} U_{\textrm{int}}^2) Q_{\textrm{bulk}}^2$. Note that $U_\textrm{bulk}-\frac{2\pi\nu L}{\hbar v_\textrm{p}} U_{\textrm{int}}^2 > 0$. 
The main interference signal $I_{\textrm{D}_3}^{\textrm{int}}$ and the reference signal $I^{\textrm{int}}_{\textrm{Ref, D}_3}$ are computed, by taking ensemble average over the thermal fluctuations of $N_L$ (see Supplementary Note 5).


\vspace{.1cm}
\noindent
{\bf Acknowledgements}. H.-S.S thanks Eddy Ardonne, Hyungkook Choi, Yunchul Chung, Dmitri Feldman, Woowon Kang, Kirill Shtengel, and Joost Slingerland for useful discussion. We thank the support by Korea NRF (Grant No. NRF-2010-00491, Grant No. NRF-2013R1A2A2A01007327; HSS), by KAIST-HRHRP (CH), and by DFG (Grant No. RO 2247/8-1; YG).

\vspace{.1cm}
\noindent
{\bf Author contributions}. C. H. has conceived the concept of topological vacuum bubbles, performed the detailed calculation, analyzed the Coulomb dominated regime, and wrote the paper. J. P. has performed the analysis, and wrote the paper. Y. G. has conceived the interferometry setup, and wrote the paper. H.-S. S. has conceived the concept of topological vacuum bubbles and the interferometry setup, analyzed the Coulomb dominated regime, wrote the paper, and supervised the project.

\clearpage

\end{document}


\section{Supplementary Information}
\centerline{\includegraphics[width=0.8\textwidth]{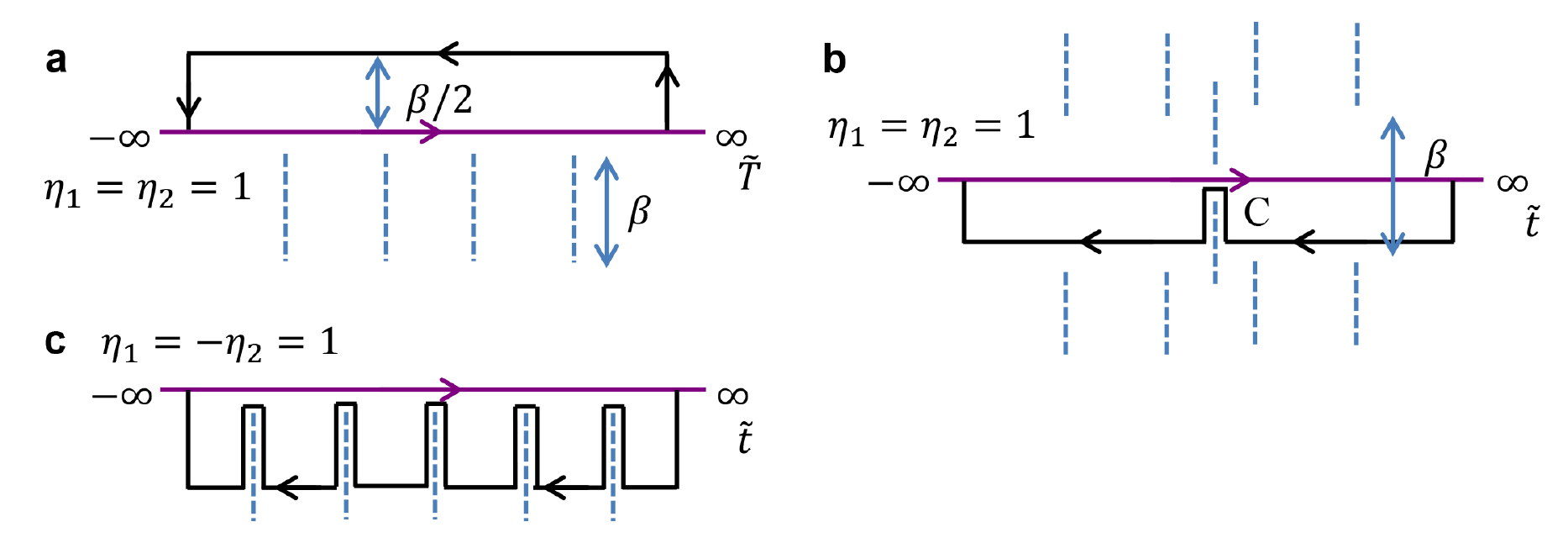}}
\noindent
{\bf Supplementary Figure~1} Integration contour for the terms of $\eta_1 = \eta_2 = 1$ and $\eta_1 = - \eta_2 = 1$ in Supplementary Equation~\ref{Current_Total}.
Magenta lines represent the real-axis integral line, while the black line a proper contour for contour integral in the complex plane of $t$. Dashed blue lines show branch-cut lines.
\label{branch1}

\vspace{.3cm}

\centerline{\includegraphics[width=0.63\textwidth]{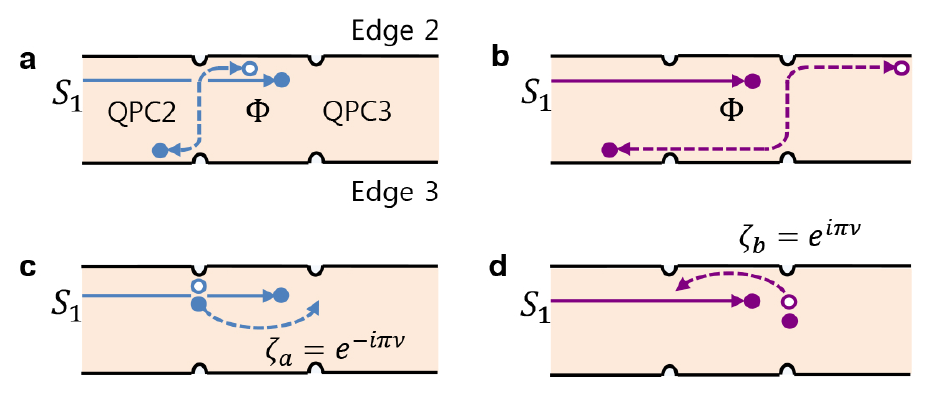}}
\noindent
{\bf Supplementary Figure~2} 
Braiding phase of Type II process. ({\bf a}) and ({\bf b}), The two interfering paths of the Type II process in Fig.~2. ({\bf c}) and ({\bf d}), Paths equivalent, respectively, to those in ({\bf a}) and ({\bf b}) in the gauge chosen in parallel to the extended edge scheme in Fig.~4 of the main text. 
In the interference of the two paths ({\bf a}) and ({\bf b}), a thermal anyon (whose trajectory is represented by dashed lines) excited at QPC2 or QPC3 {\it effectively} winds around a voltage-biased anyon (solid line) injected from Edge 1 via QPC1. The effective winding results in 
the braiding phase factor $\zeta_a^* \zeta_b = e^{i 2 \pi \nu}$. The phase factors $\zeta_a$ and $\zeta_b$ are derived from the commutation rules in Eqs.~(11) and (12). For the details, see the Supplementary Note 6.
\label{vacbub} 

\vspace{.5cm}

\paragraph*{\bf Supplementary Note 1 : Green's function.}
We provide the Green's function of bosonic fields $\phi_i$. It is defined as
\begin{equation}
G(t,x)=\langle\phi_i(t,x)\phi_i(0,0)\rangle-\langle\phi_i(0,0)\phi_i(0,0)\rangle,
\end{equation}
where $\langle \cdots \rangle$ is the ensemble average under the bare Hamiltonian $H_{\textrm{edge}, i}$.
The expression of $G$ is known~\cite{Delft} 
as $G_\beta (t,x)=-\nu\ln\left[\frac{\sin\left( \pi(\tau_0+i(t-x))/\beta\right)}{\pi\tau_0/\beta}\right]$, where $\hbar / \tau_0$ is the infrared energy cutoff and we use $v_\textrm{p} \equiv 1$ and $\beta = (k_B T)^{-1}$.
At zero temperature, $G (x,t) = -\nu\ln (\tau_0+i(t-x) / \tau_0 )$.

To describe the non-equilibrium situation by the voltage $V$ at edge 1, we use the Green's function
%
on the standard Keldysh contour with
Keldysh ordering $T_K$. 
The contour has two branches, the upper one denoted by Keldysh index $\eta = 1$ and the lower of $\eta = -1$. 
For two operators $A$ and $B$ on the upper branch, 
$T_K (AB) = AB$ if $B$ is the earlier event than $A$, and $T_K (AB) = BA$ otherwise; it is correct not to attach the phase factor for the exchange statistics in the definition of Keldysh ordering~\cite{Kim06}. 
For $A$ and $B$ on the lower, 
$T_K (AB) = BA$ if $B$ is earlier than $A$, and $T_K (AB) = AB$ otherwise. 
For $A$ on the upper and $B$ on the lower, 
$T_K (AB) = BA$.
For $A$ on the lower and $B$ on the upper, $T_K (AB) = AB$.
The Keldysh Green's function $G_{\eta\eta_0} (t,x) \equiv \langle T_K [\phi_i(t^{\eta},x)\phi_i(0^{\eta_0},0)] \rangle-\langle \phi_i(0,0)\phi_i(0,0) \rangle$ is obtained as~\cite{Kim06}
\begin{eqnarray}
&&G_{\eta\eta_0} (t,x)  = -\nu\ln \frac{\sin\left( \pi (\tau_0+i\chi_{\eta\eta_0}(t)(t-x)\right) / \beta)}{\pi\tau_0 / \beta}, \label{KeldyshGreenfunction} 
\end{eqnarray}
where $\chi_{\eta_1\eta_2}(t_1-t_2)=[ (\eta_1+\eta_2)/2 ] \textrm{sgn}(t_1-t_2)- (\eta_1-\eta_2)/2$ and $\eta_0$ and $\eta$ are Keldysh indexes.

\vspace{1cm}

\paragraph*{\bf Supplementary Note 2 : Perturbation theory.}
We compute the charge current $I_{\textrm{D}_3}$ at drain D$_3$ by the voltage $V$ at edge~1 up to order $\gamma_1^2 \gamma_2 \gamma_3$, employing perturbation theory and the Keldysh Green function in Supplementary Equation~\ref{KeldyshGreenfunction}. Hereafter, we occasionally use $\omega_0 \equiv e^* V$, $\beta \equiv (k_B T)^{-1}$, $v_\textrm{p} \equiv 1$, and $\hbar \equiv 1$.

The current $I_{\textrm{D}_3}$ is expressed as $I_{D_3} = i e^* [N_3, H] = i e^* (\langle T_2^\downarrow \rangle-\langle T_2^\uparrow \rangle +\langle T_3^\downarrow \rangle-\langle T_3^\uparrow \rangle)$.
$I_{\textrm{D}_3}$ is decomposed,
$I_{\textrm{D}_3} = I_{\textrm{D}_3}^{\textrm{dir}} + I_{\textrm{D}_3}^{\textrm{int}}$, into direct current $I_{\textrm{D}_3}^{\textrm{dir}} \propto \gamma_1^2 \gamma_2^2, \, \gamma_1^2 \gamma_3^2$, and interference current $I_{\textrm{D}_3}^{\textrm{int}} \propto \gamma_1^2 \gamma_2 \gamma_3$ depending on $\Phi$ (the leading-order contribution). 

We focus on the contribution of $\langle T_2^\downarrow \rangle$ onto $I_{\textrm{D}_3}^{\textrm{int}} \propto \gamma_1^2 \gamma_2 \gamma_3$, which is denoted by $\langle T_2^{\downarrow, \textrm{int}} \rangle$. It is 
\begin{equation}
\langle T_2^{\downarrow, \textrm{int}} \rangle  =\sum_{\eta,\eta_0\eta_1\eta_2 = \pm 1}\eta\eta_1\eta_2\int_{-\infty}^{\infty}dt_1 \int_{-\infty}^{\infty}dt_2 \int_{-\infty}^{\infty}dt \langle T_K[ T_2^{\downarrow}(0^{\eta_0})T_1^\rightarrow(t_1^{\eta_1})T_1^{\leftarrow}(t_2^{\eta_2})T_3^{\uparrow}(t^{\eta})]\rangle. \nonumber \\
\end{equation}
$\langle T_K[ T_2^{\downarrow}(0^{\eta_0})T_1^\rightarrow(t_1^{\eta_1})T_1^{\leftarrow}(t_2^{\eta_2})T_3^{\uparrow}(t^{\eta})]\rangle / \gamma_1^2 \gamma_2 \gamma_3$ is decomposed into two parts,
\begin{align}
&\langle T_K[\Psi^{\dagger}_3(0^{\eta_0},L) \Psi_2(0^{\eta_0},d)\Psi^{\dagger}_2(t_1^{\eta_1},0)\Psi_1(t_1^{\eta_1},0)\Psi_1^{\dagger}(t_2^{\eta_2},0)\Psi_2(t_2^{\eta_2},0)\Psi_2^{\dagger}(t^{\eta},d+L)\Psi_3(t^{\eta},0)]\rangle \nonumber \\
=&\langle T_K[ e^{-i\phi_3(0^{\eta_0},L)}e^{i\phi_2(0^{\eta_0},d)}e^{-i\phi_2(t_1^{\eta_1},0)}e^{i\phi_1(t_1^{\eta_1},0)}e^{-i\phi_1(t_2^{\eta_2},0)} e^{i\phi_2(t_2^{\eta_2},0)}e^{-i\phi_2(t^{\eta},d+L)}e^{i\phi_3(t^{\eta},0)} ]\rangle \nonumber \\
& \times \langle T_K[ F_3^\dagger(0^{\eta_0}) F_2(0^{\eta_0}) F_2^\dagger(t_1^{\eta_1}) F_1(t_1^{\eta_1}) F_1^\dagger(t_2^{\eta_2}) F_2(t_2^{\eta_2}) F_2^\dagger(t^{\eta}) F_3(t^{\eta})]\rangle. \nonumber 
\end{align}
We compute the former part,
using the identity (valid when the Hamiltonian of $B$'s is quadratic) for bosonic operators $B$'s, 
$\langle T_K [e^{B_1(t_1)}e^{B_2(t_2)}...e^{B_n(t_n)}] \rangle =e^{\frac{1}{2}\sum_{j=1}^{n}\langle T_K [B_j(t_j)^2]\rangle +\sum_{i<j}\langle T_K [B_i(t_i) B_j(t_j)] \rangle}$,
\begin{align}
&\langle T_K[ e^{-i\phi_3(0^{\eta_0},L)}e^{i\phi_2(0^{\eta_0},d)}e^{-i\phi_2(t_1^{\eta_1},0)}e^{i\phi_1(t_1^{\eta_1},0)} e^{-i\phi_1(t_2^{\eta_2},0)} e^{i\phi_2(t_2^{\eta_2},0)}e^{-i\phi_2(t^{\eta},d+L)}e^{i\phi_3(t^{\eta},0)} ]\rangle\nonumber\\
&=e^{2G_{\eta_1\eta_2}(t_1-t_2,0)+G_{\eta_1 \eta_0}(t_1,-d)-G_{\eta_1\eta}(t_1-t,-d-L)+G_{\eta_2\eta}(t_2-t,-d-L)-G_{\eta_2\eta_0}(t_2,-d)+G_{\eta\eta_0}(t,-L)+G_{\eta\eta_0}(t,L)}, \nonumber
\end{align}
in which the Keldysh Green's functions $G_{\eta \eta_0}$ in Supplementary Equation~\ref{KeldyshGreenfunction} appear.
The latter Klein-factor part is 
\begin{align}
& \langle T_K[ F_3^\dagger(0^{\eta_0}) F_2(0^{\eta_0}) F_2^\dagger(t_1^{\eta_1}) F_1(t_1^{\eta_1}) F_1^\dagger(t_2^{\eta_2}) F_2(t_2^{\eta_2}) F_2^\dagger(t^{\eta}) F_3(t^{\eta})]\rangle \nonumber \\
&\quad\quad\quad\quad\quad= e^{-i \omega_0 (t_1 - t_2)}
\frac{\exp[{i\pi\nu \{ \chi_{\eta_1\eta_0}(t_1) +\chi_{\eta_2\eta}(t_2-t)} \} /2]}{\exp[{i\pi\nu \{ \chi_{\eta_2\eta_0}(t_2) +\chi_{\eta_1\eta}(t_1-t)} \}/2 ]}. \nonumber
\end{align}
It is obtained, by using $F_1(t)=F_1(0) e^{-ie^* V t / \hbar}$, $F_2 (t) = F_2 (0)$, $F_3 (t) = F_3 (0)$, $F_i^\dagger F_i =F_i F_i^\dagger =1$, and the exchange statistics rule in Eq.~(12).
The final expression is 
\begin{align}
\langle T_2^{\downarrow, \textrm{int}} \rangle &=  \gamma_1^2\gamma_2\gamma_3\sum_{\eta_1,\eta_2, \eta = \pm 1} \int_{-\infty}^{0}dt\int_{-\infty}^{\infty}dt_2\int_{-\infty}^{\infty}dt_1  \frac{\eta_1\eta_2\eta e^{-i2\pi\Phi/\Phi_0^*}e^{-i\omega_0(t_1-t_2)}}{\left(\frac{\beta}{\pi}\sin(\frac{\pi}{\beta}(\tau_0+i\chi_{\eta_1\eta_2}(t_1-t_2)(t_1-t_2)))\right)^{2\nu}}\nonumber\\
&\times \frac{1}{\left(\frac{\beta}{\pi}\sin(\frac{\pi}{\beta}(\tau_0-i\eta(t+L)))\right)^{\nu}\left(\frac{\beta}{\pi}\sin(\frac{\pi}{\beta}(\tau_0-i\eta(t-L)))\right)^{\nu}}\nonumber\\
& \times \frac{\left(\sin(\frac{\pi}{\beta}(\tau_0-i\eta_1(t_1-t+d+L)))\right)^{\nu}\left(\sin(\frac{\pi}{\beta}(\tau_0-i\eta_2(t_2+d)))\right)^{\nu}}{\left(\sin(\frac{\pi}{\beta}(\tau_0-i\eta_1(t_1+d)))\right)^{\nu}\left(\sin(\frac{\pi}{\beta}(\tau_0-i\eta_2(t_2-t+d+L))\right)^{\nu}}. \nonumber
\end{align}
Computing the other terms in the same way, we obtain
\begin{align}
I_{D_3}^\textrm{int}  & =  \alpha \gamma_1^2\gamma_2\gamma_3\sum_{\epsilon\eta_1,\eta_2,\eta = \pm 1} \int_{-\infty}^{\infty}dt\int_{-\infty}^{\infty}dt_2\int_{-\infty}^{\infty}dt_1  \frac{ \epsilon \eta_1 \eta_2 \eta  e^{-i\epsilon 2\pi\Phi/\Phi_0^*}e^{-i\omega_0(t_1-t_2)}}{\left(\frac{\beta}{\pi}\sin(\frac{\pi}{\beta}(\tau_0+i\chi_{\eta_1\eta_2}(t_1-t_2)(t_1-t_2)))\right)^{2\nu}}\nonumber\\
&\times \frac{1}{\left(\frac{\beta}{\pi}\sin(\frac{\pi}{\beta}(\tau_0-i\eta(t+L)))\right)^{\nu}\left(\frac{\beta}{\pi}\sin(\frac{\pi}{\beta}(\tau_0-i\eta(t-L)))\right)^{\nu}}\nonumber\\
&\times \left[\frac{\left(\sin(\frac{\pi}{\beta}(\tau_0-i\eta_1(t_1-t+d+L)))\right)^{\nu}\left(\sin(\frac{\pi}{\beta}(\tau_0-i\eta_2(t_2+d)))\right)^{\nu}}{\left(\sin(\frac{\pi}{\beta}(\tau_0-i\eta_1(t_1+d)))\right)^{\nu}\left(\sin(\frac{\pi}{\beta}(\tau_0-i\eta_2(t_2-t+d+L))\right)^{\nu}}\right]^\epsilon, \label{Current_Total} 
\end{align}
where $\alpha \equiv (e^* / \hbar) a^{4\nu} / [ (\hbar v_\textrm{p})^{4\nu} (2 \pi a)^4]$.
Supplementary Equation~\ref{Current_Total} is written as $I_{D_3}^\textrm{int} \propto \cos (2 \pi \frac{\Phi}{ \Phi^*_0} + \theta)$ in Eq.~(5),
\begin{align}
\theta  & = \arg [  \sum_{\eta_1,\eta_2,\eta = \pm 1} \int_{-\infty}^{\infty}dt\int_{-\infty}^{\infty}dt_2\int_{-\infty}^{\infty}dt_1  \frac{ \eta_1 \eta_2 \eta   e^{-i\omega_0(t_1-t_2)}}{\left(\frac{\beta}{\pi}\sin(\frac{\pi}{\beta}(\tau_0+i\chi_{\eta_1\eta_2}(t_1-t_2)(t_1-t_2)))\right)^{2\nu}}\frac{1}{\left(\frac{\beta}{\pi}\sin(\frac{\pi}{\beta}(\tau_0-i\eta(t+L)))\right)^{\nu}}\nonumber\\
&\times \frac{1}{\left(\frac{\beta}{\pi}\sin(\frac{\pi}{\beta}(\tau_0-i\eta(t-L)))\right)^{\nu}}
\frac{\left(\sin(\frac{\pi}{\beta}(\tau_0-i\eta_1(t_1-t+d+L)))\right)^{\nu}\left(\sin(\frac{\pi}{\beta}(\tau_0-i\eta_2(t_2+d)))\right)^{\nu}}{\left(\sin(\frac{\pi}{\beta}(\tau_0-i\eta_1(t_1+d)))\right)^{\nu}\left(\sin(\frac{\pi}{\beta}(\tau_0-i\eta_2(t_2-t+d+L))\right)^{\nu}}   ]. \label{Current_Total_phase}
\end{align}

We omit the expression of the direct-current part $I_{\textrm{D}_3}^{\textrm{dir}}$ of $I_{\textrm{D}_3}$ (of the order of $\gamma_1^2 \gamma_2^2$ and $\gamma_1^2 \gamma_3^2$). The whole expression of  $I_{\textrm{D}_3}$ (up to the order of $\gamma_1^2 \gamma_2 \gamma_3$) is zero at $V = 0$, as expected. 

\vspace{1cm}

\paragraph*{\bf Supplementary Note 3 : Semiclassical Approximation.}
We derive Eqs.~(3), (4) and~(6) in the main text, by performing the integrals in Supplementary Equation~\ref{Current_Total} 
(or those in Supplementary Equation~\ref{Current_Total_phase})
in the semiclassical regime of $e^* V \gg k_B T \gg \hbar v_\textrm{p}/L$. 

In the semiclassical regime, the wave packet size of an injected anyon from S$_1$ is $W \sim \hbar v_\textrm{p} / (e^* V) \ll L$, hence, the anyon is thought of as particle-like (rather than wave-like).
And the injection of hole-like anyons (induced by thermal fluctuations) from S$_1$ can be ignored. Thermally excited virtual anyons also have size much smaller than $L$. Then, the braiding of the anyons results in the well-defined exchange statistics phase of $\pi \nu$. 

In Supplementary Equation~\ref{Current_Total}, there are 16 configurations of $\eta_1$, $\eta_2$, $\eta$, and $\epsilon$. 
$\eta_1$ and $\eta_2$ $(=\pm1)$ are Keldysh indexes for tunneling at QPC1, while $\eta$ is for tunneling at QPC2 or QPC3. $\epsilon = \pm 1$ represents the winding direction of anyons around Aharonov-Bohm flux $\Phi$. The terms of Supplementary Equation~\ref{Current_Total} are classified, depending on whether $\eta_1 = \eta_2$.
In the regime of $e^* V \gg k_B T \gg \hbar v_\textrm{p}/L$, some terms of $\eta_1 \ne \eta_2$ are represented by connected Feynman diagrams, and contribute to Type I-1 and Type I-2.
The other terms of $\eta_1 \ne \eta_2$ and all the terms of $\eta_1 = \eta_2$ are represented by vacuum bubbles.

Some of the vacuum bubbles of $\eta_1 \ne \eta_2$ describe the virtual anyonic excitations winding around a real anyon injected from S$_1$ (similarly to Fig.~1a in the main text).
Each of such bubble diagrams has a partner diagram among the terms of $\eta_1 = \eta_2$,
in which a vacuum bubble does not wind around the real anyon (similarly to Fig.~1b in the main text); each of the diagrams of $\eta_1 = \eta_2$ is decomposed into two disconnected and independent subdiagrams, one describing the real anyon, and the other for the virtual anyonic excitations not winding around the real anyon. 
A bubble diagram of $\eta_1 \ne \eta_2$ winding around a real anyon, and its partner diagram
of $\eta_1 = \eta_2$ cancel each other. The cancellation is only partial, resulting in
the interference current of Type~II and Type~III in Eqs.~(3) and (4), and disobeying the linked cluster theorem; the cancellation is exact in the case of bosons or fermions.
Note that there are the terms of $\eta_1 \ne \eta_2$ not involving any braiding effect between real anyons and virtual anyons. 
These terms show divergence, but are fully canceled by their partner diagrams of $\eta_1 = \eta_2$,  
 satisfying the linked cluster theorem.

We now further compute Supplementary Equation~\ref{Current_Total} in the limit of $e^* V \gg k_B T \gg \hbar v_\textrm{p}/L$.
For example, 
\begin{align}
& \textrm{the} \textrm{ term of } (\eta_1=\eta_2= \pm 1, \epsilon=1) \textrm{ in Supplementary Equation~\ref{Current_Total}} = 2\alpha \gamma_1^2\gamma_2\gamma_3 \sum_{\eta,\eta_1}\eta \int_{-\infty}^{\infty}dt\int_{-\infty}^{\infty}d\tilde{t} \int_{-\infty}^{\infty}d\tilde{T}\nonumber \\
& \times  \frac{ e^{-i 2\pi\Phi/\Phi_0^*}e^{-i\omega_0 2\tilde{t}}}{\left(\frac{\beta}{\pi}\sin(\frac{\pi}{\beta}(\tau_0+i\eta_1\textrm{sgn}(\tilde{t})2\tilde{t})))\right)^{2\nu} \left(\frac{\beta}{\pi}\sin(\frac{\pi}{\beta}(\tau_0-i\eta(t+L)))\right)^{\nu}\left(\frac{\beta}{\pi}\sin(\frac{\pi}{\beta}(\tau_0-i\eta(t-L)))\right)^{\nu}}\nonumber\\
& \times \left[\frac{\left(\sin(\frac{\pi}{\beta}(\tau_0-i\eta_1(\tilde{T}+\tilde{t}-t+d+L)))\right)^{\nu}\left(\sin(\frac{\pi}{\beta}(\tau_0-i\eta_1 (\tilde{T}-\tilde{t}+d)))\right)^{\nu}}{\left(\sin(\frac{\pi}{\beta}(\tau_0 -i\eta_1 (\tilde{T}+\tilde{t}+d)))\right)^{\nu}\left(\sin(\frac{\pi}{\beta}(\tau_0 -i\eta_1 (\tilde{T}-\tilde{t}-t+d+L))\right)^{\nu}}\right] \nonumber \\
&\simeq 2\alpha \gamma_1^2\gamma_2\gamma_3 \sum_{\eta,\eta_1} \eta \int_{-\infty}^{\infty}dt\int_{-\infty}^{\infty}d\tilde{T}\int_C d\tilde{t}  \nonumber\\
& \times \frac{ e^{-i 2\pi\Phi/\Phi_0^*}e^{-i\omega_0 2\tilde{t}}\left[1+\frac{2\pi\nu}{\beta}\tilde{t}\textrm{ sgn}(t-L)\left\lbrace\textrm{sgn}(\tilde{T}-t+d+L)\textrm{sgn}(\tilde{T}+d)-1\right\rbrace\right]}{\left(\frac{\beta}{\pi}\sin(\frac{\pi}{\beta}(\tau_0+i \eta_1 \textrm{sgn}(\tilde{t})2\tilde{t}))\right)^{2\nu} \left(\frac{\beta}{\pi}\sin(\frac{\pi}{\beta}(\tau_0-i\eta(t+L)))\right)^{\nu}\left(\frac{\beta}{\pi}\sin(\frac{\pi}{\beta}(\tau_0-i\eta(t-L)))\right)^{\nu}}\nonumber
\end{align}
In the first equality, we put $\tilde{t}=(t_1-t_2)/2$ and $\tilde{T}=(t_1+t_2)/2$ into Supplementary Equation~\ref{Current_Total}. In the second, we change the integral range of $\tilde{T}$ from $\tilde{T} \in (-\infty, \infty)$ to $\tilde{T} \in (-\infty + i \eta_1 \beta / 2, \infty + i \eta_1 \beta / 2)$ using the contour shown in Supplementary Figure~1a, and that of $\tilde{t}$ from $\tilde{t} \in (-\infty, \infty)$ to the vertical parts $C$ near Re$(\tilde{t})=0$ of the contour in Supplementary Figure~1b.
Note that the integral of $\tilde{t}$ along the horizontal black line of Supplementary Figure~1b is negligible in the limit of $e^* V \gg k_B T \gg \hbar v_\textrm{p}/L$.
Then
\begin{align}
&\textrm{Sum of the} \,\, (\eta_1= \eta_2= 1, \epsilon=1) \,\, \textrm{term and the} \,\, (\eta_1= \eta_2= -1, \epsilon=1) \,\, \textrm{term} \nonumber \\
&\simeq \alpha \gamma_1^2\gamma_2\gamma_3   \int_{-\infty}^{\infty}dt \int_{-\infty}^{\infty}d\tilde{T} \left[1 + \frac{2\pi i \nu(1-2\nu) \textrm{ sgn}(t-L)}{\omega\beta}\left\lbrace\textrm{sgn}(\tilde{T}-t+d+L)\textrm{sgn}(\tilde{T}+d)-1\right\rbrace\right]\nonumber
\\
&\quad\qquad\qquad\times  \sum_{\eta} \eta \frac{A_1   e^{-i 2\pi\Phi/\Phi_0^*}}{\left(\frac{\beta}{\pi}\sin(\frac{\pi}{\beta}(\tau_0-i\eta(t+L)))\right)^{\nu}\left(\frac{\beta}{\pi}\sin(\frac{\pi}{\beta}(\tau_0-i\eta(t-L)))\right)^{\nu}}
, \label{sum_term}
\end{align}
where $A_1=\int_{-\infty}^{\infty} d \tilde{t} e^{-i\omega_0 \tilde{t}} \left(\frac{\beta}{\pi}\sin(\frac{\pi}{\beta}(\tau_0 -i \tilde{t})\right)^{-2\nu}\simeq \frac{2\pi}{\Gamma(2\nu)}\omega_0^{2\nu-1}$ and $\Gamma(2 \nu)$ is the Gamma function.
The divergence of Supplementary Equation~\ref{sum_term} is fully canceled by some terms of $\eta_1 \ne \eta_2$. 
Next, we consider 
\begin{align}
\textrm{the} & \textrm{ term of } (\eta_1= 1,\eta_2= -1, \epsilon=1)  = 2\alpha \gamma_1^2\gamma_2\gamma_3 \sum_{\eta} (-\eta)  \int_{-\infty}^{\infty}dt\int_{-\infty}^{\infty}d\tilde{T}\int_{-\infty}^{\infty}d\tilde{t}\nonumber \\& \times \frac{ e^{-i 2\pi\Phi/\Phi_0^*}e^{-i\omega_0 2\tilde{t}}}{\left(\frac{\beta}{\pi}\sin(\frac{\pi}{\beta}(\tau_0-i 2\tilde{t}))\right)^{2\nu} \left(\frac{\beta}{\pi}\sin(\frac{\pi}{\beta}(\tau_0-i\eta(t+L)))\right)^{\nu}\left(\frac{\beta}{\pi}\sin(\frac{\pi}{\beta}(\tau_0-i\eta(t-L)))\right)^{\nu}}\nonumber\\
& \times \left[\frac{\left(\sin(\frac{\pi}{\beta}(\tau_0-i(\tilde{T}+\tilde{t}-t+d+L)))\right)^{\nu}\left(\sin(\frac{\pi}{\beta}(\tau_0+i(\tilde{T}-\tilde{t}+d)))\right)^{\nu}}{\left(\sin(\frac{\pi}{\beta}(\tau_0-i(\tilde{T}+\tilde{t}+d)))\right)^{\nu}\left(\sin(\frac{\pi}{\beta}(\tau_0+i(\tilde{T}-\tilde{t}-t+d+L))\right)^{\nu}}\right] \nonumber \\
& \simeq \alpha \gamma_1^2\gamma_2\gamma_3  \int_{-\infty}^{\infty}dt\left(-A_2^2 e^{i\pi\nu\textrm{sgn}(t-L)}-A_2^2 e^{i\omega_0 (t-L)} - A_1 \int_{-\infty}^{\infty}d\tilde{T} e^{-i\pi\nu(\textrm{sgn}(\tilde{T}-t+d+L)-\textrm{sgn}(\tilde{T}+d))}\right.\nonumber\\
&\left.\times\left[1 + \frac{2\pi i\nu(1-2\nu)\textrm{ sgn}(t-L)}{\omega\beta}\left\lbrace\textrm{sgn}(\tilde{T}-t+d+L)\textrm{sgn}(\tilde{T}+d)-1\right\rbrace\right]  \right)\nonumber\\
& \times \sum_\eta \eta \frac{e^{-i 2\pi\Phi/\Phi_0^*}}{\left(\frac{\beta}{\pi}\sin(\frac{\pi}{\beta}(\tau_0-i\eta(t+L)))\right)^{\nu}\left(\frac{\beta}{\pi}\sin(\frac{\pi}{\beta}(\tau_0-i\eta(t-L)))\right)^{\nu}}. \label{typeII-contribution}
\end{align}
Here $A_2=\int_{-\infty}^{\infty} dt e^{-i\omega_0 t}  \left(\frac{\beta}{\pi}\sin(\frac{\pi}{\beta}(\tau-it)) \right)^{-\nu} \simeq \frac{2\pi}{\Gamma(\nu)}\omega_0^{\nu-1}$, $\tilde{t}=(t_1-t_2)/2$, and $\tilde{T}=(t_1+t_2)/2$.
The integral over $\tilde{t}$ is done in the regime of $e^* V \gg k_B T$, using the contour in Supplementary Figure~1c.

The terms of $(\eta_1= 1,\eta_2= -1, \epsilon=1)$ with coefficient $A_2$ in Supplementary Equation~\ref{typeII-contribution} describe Types I-1 and I-2, which have no counterpart in the terms of $\eta_1 = \eta_2$. The divergence of some terms with coefficient $A_1$ in Supplementary Equation~\ref{typeII-contribution} is fully canceled by the corresponding terms of $\eta_1 = \eta_2$. The remaining terms with $A_1$ are partially canceled by the corresponding terms of $\eta_1 = \eta_2$, describing Types II and III.
Indeed, the difference of the terms with $A_1$ between Supplementary Equations~\ref{sum_term} and~\ref{typeII-contribution} is proportional to
\begin{align}
&- A_1 \int_{-\infty}^{\infty}d\tilde{T} \left(e^{-i\pi\nu(\textrm{sgn}(\tilde{T}-t+d+L)-\textrm{sgn}(\tilde{T}+d))}-1\right)\nonumber\\
&\quad\times\left[1 + \frac{2\pi i \nu(1-2\nu)\textrm{ sgn}(t-L)}{\omega\beta}\left\lbrace\textrm{sgn}(\tilde{T}-t+d+L)\textrm{sgn}(\tilde{T}+d)-1\right\rbrace\right]\nonumber\\
&\quad=-2i A_1 (t-L) e^{i\pi\nu\textrm{sgn}(t-L)}   \left[1+\frac{2\pi i\nu(1-2\nu)\textrm{ sgn}(t-L)}{\omega\beta}\right] \sin\pi\nu. \nonumber
\end{align}
The difference is proportional to $\sin \pi \nu$, and describes the topological vacuum bubbles corresponding to the braiding discussed in Fig.~1 and Eqs.~(3) and (4).
The term of $\eta_1= - 1$, $\eta_2= 1$, and $\epsilon=1$ describes the injection of hole-like anyons from edge 1 to edge 2, and is negligible in our regime of $e^* V \gg k_B T$. 
The terms with $\epsilon = -1$ are obtained in the same way, as they are for the processes of winding $\Phi$ in the opposite direction to those of $\epsilon = 1$.


Now we collect all the terms of $I_{D_3}^\textrm{int}$,
\begin{align}
& I_{D_3}^\textrm{int} = \,\alpha \gamma_1^2\gamma_2\gamma_3 \sum_{\eta} (-\eta) \int_{-\infty}^{\infty}dt \frac{1}{\left( \frac{\beta}{\pi}\sin(\frac{\beta}{\pi}(\tau_0-i\eta(t+L))) \right)^{\nu} \left( \frac{\beta}{\pi}\sin(\frac{\pi}{\beta}(\tau_0-i\eta(t-L))) \right)^{\nu}}\nonumber
\\ \times & [(-2i\sin(\pi\nu)A_1 \lbrace 1+f(t)\rbrace e^{i\pi\nu\textrm{sgn}(t-L)}(t-L)-A_2^2 e^{i\pi\nu\textrm{sgn}(t-L)} -A_2^2 e^{i\omega_0(t-L)})e^{-i2\pi\Phi/\Phi_0^*} \nonumber
\\&- (2i\sin(\pi\nu)A_1 \lbrace 1-f(t)\rbrace e^{-i\pi\nu\textrm{sgn}(t-L)}(t-L)-A_2^2 e^{-i\pi\nu\textrm{sgn}(t-L)} -A_2^2 e^{-i\omega_0(t-L)})e^{i2\pi\Phi/\Phi_0^*}], \label{first_approx}
\end{align}
where $f(t)=\frac{2\pi i \nu(1-2\nu)\textrm{ sgn}(t-L)}{\omega\beta}$.  The integral of $t$ is further computed for $k_B T \gg \hbar v_\textrm{p}/L$;  we use $\frac{\beta}{\pi}\sin(\frac{\pi}{\beta}(\tau_0-it))
\simeq -i\textrm{sgn}(t)\frac{\beta}{2\pi}e^{\frac{\pi}{\beta}|t|}$ for $t\gg \beta$, 
$\int_{-\infty}^{0} dt e^{\pi\nu t / \beta} \left(\frac{\beta}{\pi}\sin(\frac{\beta}{\pi}(\tau_0+i t))\right)^{-\nu}= (\frac{2\pi}{\beta})^\nu \frac{\beta}{2\pi}e^{i \pi\nu /2} \pi\csc(\pi\nu)$,
and
$\int_{-\infty}^{0} dt e^{\pi\nu t / \beta} t \left(\frac{\beta}{\pi}\sin(\frac{\beta}{\pi}(\tau_0+i t))\right)^{-\nu} = (\frac{2\pi}{\beta})^\nu (\frac{\beta}{2\pi})^2 \pi e^{i \pi\nu / 2} (\psi(1)-\psi(\nu)) \csc(\pi\nu)$, 
where $\psi(x)$ is the polygamma function. The integration leads to Eq.~(3),
\begin{align}
I_{\textrm{D}_3}^{\textrm{I-1}} \,=&\, \frac{16\pi^3(2\pi)^\nu \alpha}{\Gamma(\nu)^3}
\gamma_1^2 \gamma_2 \gamma_3 (e^*V)^{3\nu-3} (k_B T)^\nu \cos(e^*V L / \hbar v_\textrm{p} -\pi\nu/2) \nonumber \\
&  \times e^{-2L/L_T} \cos(2\pi\Phi/\Phi_0^*+e^* V L/\hbar v_\textrm{p}), \nonumber \\ 
I_{\textrm{D}_3}^{\textrm{I-2}} \,=&\, - \frac{8\pi(2\pi)^{2\nu} \alpha}{\nu\Gamma(\nu)^2} \gamma_1^2 \gamma_2 \gamma_3 (e^*V)^{2\nu-2}(k_B T)^{2\nu-1}  e^{-2L/L_T} \sin^2\pi\nu \cos(2\pi\Phi/\Phi_0^*), \nonumber \\ 
I_{\textrm{D}_3}^{\textrm{II}}\, =&\, \frac{4(2\pi)^{2\nu} \pi \alpha  \csc (\pi\nu)}{\Gamma(2\nu)} \gamma_1^2 \gamma_2 \gamma_3 (e^* V)^{2\nu-1}(k_B T)^{2\nu-1} \sqrt{1+h_1(T/V)}
\frac{2L + C L_T}{\hbar v_\textrm{p}} \nonumber \\
& \times e^{-2L/L_T} \sin^2\pi\nu \cos(2\pi\Phi/\Phi_0^{*}+\pi\nu + h_2(T/V)), \nonumber \\ 
I_{\textrm{D}_3}^{\textrm{III}}\, =&\, \frac{4(2\pi)^{2\nu} \pi \alpha  \csc (\pi\nu)}{\Gamma(2\nu)} \gamma_1^2 \gamma_2 \gamma_3 (e^* V)^{2\nu-1} (k_B T)^{2\nu-1} \sqrt{1+ h_1(T/V)}
\frac{C L_T}{\hbar v_\textrm{p}}  \nonumber\\
&\times e^{-2L/L_T} \sin^2\pi\nu \cos(2\pi\Phi/\Phi_0^{*}-\pi\nu - h_2(T/V)), \nonumber
\end{align}  
where $C \simeq \nu [\psi (\nu) - \psi (1)]/2$, $\alpha \equiv (e^* / \hbar) a^{4\nu} / [ (\hbar v_\textrm{p})^{4\nu} (2 \pi a)^4]$, and $a$ is the short-length cutoff;
$C = 0.43$ for $\nu =1/3$ and $C \to 0.5$ as $\nu \to 0$.
Note that $f(t) \propto k_B T / \omega$ in Supplementary Equation~\ref{first_approx} leads to the higher-order corrections of $h_1 (T/V) \equiv \left(\frac{2\pi\nu k_B T}{e^* V}\right)^2$ and
$h_2 (T/V) \equiv (1-2\nu)\arctan(\frac{2\pi\nu k_B T}{e^* V})$. These corrections are ignored in Eqs.~(3) and (4), considering the limit $e^* V \gg k_B T$, while they are included in Eq.~(6) for better comparison with the numerical result in Fig.~3. 

\vspace{2cm}

\paragraph*{\bf Supplementary Note 4 : Reference signal.} 
In the main text, we suggest to measure a reference interference signal $I^{\textrm{int}}_{\textrm{Ref, D}_3}$, by applying infinitesimal voltage $V_\textrm{ref}/2$ at source S$_2$ and $-V_{\textrm{ref}}/2$ at S$_3$, but turning off the voltage $V$ at S$_1$.  It is obtained~\cite{Chamon} up to the lowest order in $\gamma$'s,
\begin{align}
I^{\textrm{int}}_{\textrm{Ref, D}_3} 
&\simeq \frac{e^*}{\hbar}\frac{\gamma_2 \gamma_3 \,a^{2\nu}}{(\hbar v_\textrm{p})^{2\nu}(2\pi a)^2} e^* V_{\textrm{ref}}  (k_B T)^{2\nu-2} e^{-2L/L_T}\cos \frac{2\pi\Phi}{\Phi^*_0} \quad \quad \textrm{for} \,\,  k_B T\gg e^* V_\textrm{ref}. \label{reference_result}
\end{align}

Since the reference signal $I^{\textrm{int}}_{\textrm{Ref, D}_3}$ and the main interference signal $I^{\textrm{int}}_{\textrm{D}_3}$ are obtained from the same setup under the same external parameters (temperature, gate voltage, magnetic field, etc), the possible side effects affecting the main signal, including the external-parameter dependence of the size, shape, QPC tunneling, and anyon excitations of the interferometry, appear in the reference in the same manner.
Therefore, by comparing the reference with the main signal, one can exclude the side effects and  unambiguously detect the interference phase shift $\theta$ in Eq.~(6), hence, the fractional statistics. This strategy works well both in the pure Aharonov-Bohm regime and the Coulomb-dominated regime.



\vspace{1cm}

\paragraph*{\bf Supplementary Note 5 : Coulomb dominated regime.} We discuss about the Coulomb interaction between the edge and the bulk of the interferometry. 
There are two regimes of Fabry-Perot interferometry, the pure Aharonov-Bohm regime and the Coulomb dominated regime. In the former regime, the interaction is negligible, while it is crucial in the latter~\cite{Halperin}. 
The Fabry-Perot interferometers of recent experiments~\cite{An, Ofek, McClure, Willett,Camino} in the fractional quantum Hall regime are in the Coulomb dominated regime.
We below compute the interference current $I^{\textrm{int}}_{\textrm{D}_3}$ in the presence of the Coulomb interaction, and show that Equation~(6) of the main text is applicable to both of the pure Aharonov-Bohm regime and the Coulomb dominated regime. 
We then generalize this to the quantum Hall regime of filling factor $\nu' = \nu + \nu_0$ where $\nu = 1/ (2n+1)$ and $\nu_0$ is a nonzero integer.

We compute $I^{\textrm{int}}_{\textrm{D}_3}$, combining our chiral Luttinger liquid theory with the capacitive interaction model~\cite{Halperin} that successfully describes the Coulomb dominated regime. 
The interferometer Hamiltonian $H = \sum_i H_{\textrm{edge}, i} + H_\textrm{tun}$ is modified by the Coulomb interaction as
\begin{align}
H&\to H+U_{\textrm{int}}Q_{\textrm{bulk}} (\int_{d}^{d+L} dx:\partial_x \phi_2(x):+\int_{0}^{L} dx:\partial_x \phi_3(x): )+U_\textrm{bulk} Q_{\textrm{bulk}}^2\nonumber\\
&=\sum_{i=1,2,3}\frac{\hbar v_\textrm{p}}{4\pi\nu}\int_{-\infty}^{\infty} dx: (\partial_x \bar{\phi}_i(x))^2+ H_{\textrm{bulk}} + H_\textrm{tun}.\label{NEWH}
\end{align}
Here, $Q_{\textrm{bulk}}=\nu B A_{\textrm{area}}/\Phi_0 +\nu N_L -\bar{q}$ is the number of the net charges localized within the interferometer bulk (inside the interference loop), $A_\textrm{area}$ is the area of the interferometer, $N_L$ is the net number of quasiparticles minus quasiholes, and $\bar{q}$ is the number of positive background charges induced by the gate voltage applied to the interferometer. $U_\textrm{int}$ is the strength of Coulomb interaction between the charges of the interferometer edge and the charges localized in the interferometer bulk, and $U_\textrm{bulk}$ is the strength of interaction between the bulk charges. 
In the second equality of Supplementary Equation~\ref{NEWH}, we introduce a boson field $\bar{\phi}_i$ for each edge $i$, $\bar{\phi}_i(x)=\phi_i(x)+\frac{2\pi\nu}{\hbar v_\textrm{p}} U_\textrm{int} Q_{\textrm{bulk}} \int_{-\infty}^{x} K_i(x')dx'$,
where $K_2(x)= 1$ for $d<x<d+L$, $K_3(x) = 1$ for $0<x<L$, and $K_i(x) = 0$ otherwise.
The second term of $\bar{\phi}$ describes the charges
$-\frac{2\pi\nu}{\hbar v_\textrm{p}} U_\textrm{int} Q_{\textrm{bulk}}$ induced per unit length by the interaction.
In Supplementary Equation~\ref{NEWH}, the Hamiltonian is quadratic in $\bar{\phi}_i$ and has $H_{\textrm{bulk}}= (U_\textrm{bulk}-\frac{2\pi\nu L}{\hbar v_\textrm{p}} U_{\textrm{int}}^2) Q_{\textrm{bulk}}^2$. Note that $U_\textrm{bulk}-\frac{2\pi\nu L}{\hbar v_\textrm{p}} U_{\textrm{int}}^2 > 0$. The Keldysh Green's function of $\bar{\phi}_i$ is identical to that of $\phi_i$ in Supplementary Equation~\ref{KeldyshGreenfunction}.

%

Repeating the calculation for the case of $U_\textrm{int} = U_\textrm{bulk} = 0$ and $N_L = 0$, we compute the main interference signal 
$I^{\textrm{int}}_{\textrm{D}_3}$ and the reference signal $I^{\textrm{int}}_{\textrm{Ref, D}_3}$.
The reference signal is written as
\begin{align}
\langle I^{\textrm{int}}_{\textrm{Ref, D}_3} \rangle\propto \textrm{Re} \left[ \frac{ \sum_{N_L=-\infty}^{\infty} e^{-\beta H_{\textrm{bulk}}}
\exp (-i2\pi(\nu B A_\textrm{area}/\Phi_0 +\nu N_L)+i2\pi\frac{\nu U_{\textrm{int}} 2L}{\hbar v_\textrm{p}}Q_{\textrm{bulk}} )}{\sum_{N_L=-\infty}^{\infty}e^{-\beta H_{\textrm{bulk}}}}\right]. \label{bulkInte}
\end{align}
Here, $\exp(-i2\pi\nu N_L)$ counts the phase $2 \pi \nu N_L$ by the braiding of an interfering anyon around the anyons of the interferometer bulk, and $\sum_{N_L} \cdots$
is for the ensemble average over thermal fluctuations of $N_L$.
Substituting $Q_{\textrm{bulk}}=\nu B A_{\textrm{area}}/\Phi_0 +\nu N_L -\bar{q}$, we
obtain
\begin{align}
\langle I^{\textrm{int}}_{\textrm{Ref, D}_3} \rangle \propto & e^{\frac{-\pi^2\left(1-\frac{2\nu L}{\hbar v_\textrm{p}} U_{\textrm{int}}\right)^2}{\beta ( U_\textrm{bulk}-\frac{2\pi\nu L}{\hbar v_\textrm{p}} U_{\textrm{int}}^2)}}
\frac{\textrm{Re}\left[e^{-i2\pi\bar{q}} \Theta_3 (\frac{\pi B A_{\textrm{area}}} {\Phi_0} -\frac{\pi\bar{q}}{\nu} +\frac{i\pi^2 \left(1-\frac{2\nu L}{\hbar v_\textrm{p}} U_{\textrm{int}}\right)}{\beta\nu\left( U_\textrm{bulk} -\frac{2\pi\nu L}{\hbar v_\textrm{p}} U_{\textrm{int}}^2\right)}, \exp (\frac{-\pi^2 \beta^{-1}\nu^{-2} }{(U_\textrm{bulk}-\frac{2\pi\nu L}{\hbar v_\textrm{p}} U_{\textrm{int}}^2)} ) ) 
\right]}{\Theta_3 ( \frac{\pi B A_{\textrm{area}}}{ \Phi_0} -\frac{\pi \bar{q}}{\nu}, \exp (\frac{-\pi^2 \beta^{-1}\nu^{-2}}{(U_\textrm{bulk} -\frac{2\pi\nu L}{\hbar v_\textrm{p}}U_{\textrm{int}})} ) )}. \label{Expression_fig3_ref}
\end{align}
$\Theta_3(z,q)=1+2\sum_{n=1}^{\infty}q^{n^2}\cos(2nz)$ is the Jacobi theta function and obeys $\Theta_3(z,q)=\Theta_3(z+\pi,q)$. 
In the Coulomb dominated regime of $\hbar v_\textrm{p}/(2 \nu L)\simeq U_{\textrm{int}}$, 
$I^{\textrm{int}}_{\textrm{Ref, D}_3}$ is independent of the magnetic field $B$,
as $\Theta_3$'s in the numerator and the denominator cancel each other. 
In the pure Aharonov-Bohm regime, $I^{\textrm{int}}_{\textrm{Ref, D}_3}$ is a periodic function of $B$ with periodicity $\Phi_0/A_{\textrm{area}}$. 

On the other hand, $I^{\textrm{int}}_{\textrm{Ref, D}_3}$ shows periodic oscillations as a function of the gate voltage in both the regimes, since $\bar{q}$ and $A_\textrm{area}$ depend on the gate voltage. To see this, we write $I^{\textrm{int}}_{\textrm{Ref, D}_3}$ as 
\begin{align}
\langle I^{\textrm{int}}_{\textrm{Ref, D}_3} \rangle \propto  \frac{\textrm{Re} \left[\Theta_3 ( \left(\pi\bar{\alpha}-\frac{\pi\bar{\gamma}}{\nu}\right) V_\textrm{G} + \frac{ i\pi^2 \left(1-\frac{2\nu L}{\hbar v_\textrm{p}} U_{\textrm{int}}\right)}{\beta\nu\left( U_\textrm{bulk} -\frac{2\pi\nu L}{\hbar v_\textrm{p}} U_{\textrm{int}}^2\right)}, \exp (\frac{-\pi^2\beta^{-1}\nu^{-2}}{(U_\textrm{bulk}-\frac{2\pi\nu L}{\hbar v_\textrm{p}} U_{\textrm{int}}^2)} ) ) e^{-i2\pi\bar{\gamma} V_\textrm{G}}\right]}{\Theta_3 ( \left( \pi\bar{\alpha} -\frac{\pi\bar{\gamma}}{\nu}\right)V_\textrm{G}, \exp (-\frac{\pi^2\beta^{-1}\nu^{-2}}{(U_\textrm{bulk} -\frac{2\pi\nu L}{\hbar v_\textrm{p}}U_{\textrm{int}})} ) )}, \label{Expression_fig3_ref2}
\end{align}
where $\bar{\gamma} \equiv \frac{d\bar{q}}{dV_\textrm{G}}$ and $\bar{\alpha} \equiv \frac{B}{\Phi_0}\frac{dA_{\textrm{area}}}{dV_\textrm{G}}$.
The oscillation period is $1/\bar{\gamma}$ ($= V_{\textrm{G},0}$) in the Coulomb dominated regime of $U_\textrm{int} \simeq \hbar v_\textrm{p}/(2 \nu L)$ and also in the pure Aharonov-Bohm regime of $U_{\textrm{int}}\simeq 0$ and $\beta U_\textrm{bulk} \ll 1$.
All these results are in good agreement with the previous result in~\cite{Halperin}. 

And we obtain the main interference signal $I^{\textrm{int}}_{\textrm{D}_3}$, 
combining Supplementary Equation~\ref{Current_Total} and the ensemble average over thermal fluctuations of $N_L$ which is considered in Supplementary Equation~\ref{bulkInte},
\begin{align}
\langle I^{\textrm{int}}_{\textrm{D}_3}\rangle & \propto \frac{\textrm{Re}\left[e^{-i2\pi\bar{q}}e^{-i\theta}\Theta_3 ( \frac{\pi B A_{\textrm{area}}}{\Phi_0} -\frac{\pi\bar{q}}{\nu} +\frac{i\pi^2\left(1-\frac{2\nu L}{\hbar v_\textrm{p}} U_{\textrm{int}}\right)}{\beta\nu\left( U_\textrm{bulk} -\frac{2\pi\nu L}{\hbar v_\textrm{p}} U_{\textrm{int}}^2\right)}, \exp (\frac{-\pi^2\beta^{-1}\nu^{-2}}{ (U_\textrm{bulk}-\frac{2\pi\nu L}{\hbar v_\textrm{p}} U_{\textrm{int}}^2)} ) ) \right]}{\Theta_3 ( \pi\frac{\pi B A_{\textrm{area}}}{\Phi_0} -\frac{\pi\bar{q}}{\nu}, \exp ( -\frac{ \pi^2 \beta^{-1} \nu^{-2}} {(U_\textrm{bulk} -\frac{2\pi\nu L}{\hbar v_\textrm{p}}U_{\textrm{int}})} ) ) }, \label{Expression_fig3}
\end{align}
Notice that Supplementary Equation~\ref{Expression_fig3} is identical to Supplementary Equation.~\ref{Expression_fig3_ref} except $e^{-i\theta}$.  We find that the phase shift $\theta$ in Supplementary Equation~\ref{Expression_fig3} satisfies the expression in Supplementary Equation~\ref{Current_Total_phase}, hence, follows Equation~(6) (see the main text) in the semiclassical regime of $e^* V \gg k_B T \gtrsim \hbar v_\textrm{p}/L$.
Supplementary Equations~\ref{Expression_fig3_ref}, \ref{Expression_fig3_ref2}, and \ref{Expression_fig3} are used for drawing Fig.~3.

The phase $\theta$ results from the topological vacuum bubbles.
$\theta$ is detectable
both in the Coulomb dominated regime and the pure Aharonov-Bohm regime. In the Coulomb dominated regime of $U_\textrm{int} \simeq \hbar v_\textrm{p}/(2\nu L)$, $I^{\textrm{int}}_{\textrm{D}_3} \propto \textrm{Re} [e^{-i2\pi\bar{\gamma} V_\textrm{G}}e^{-i\theta}]$ and $I^{\textrm{int}}_{\textrm{Ref, D}_3} \propto \textrm{Re} [e^{-i2\pi\bar{\gamma} V_\textrm{G}}]$ are independent of the magnetic field. In this case, the interference pattern of $I^{\textrm{int}}_{\textrm{D}_3}$ is shifted by $\theta$ from that of $I^{\textrm{int}}_{\textrm{Ref, D}_3}$ as the gate voltage $V_\textrm{G}$ varies. 
On the other hand, in the pure Aharonov-Bohm regime of $U_{\textrm{int}}\simeq 0$ and when $\beta U_\textrm{bulk} \ll 1$, $I^{\textrm{int}}_{\textrm{D}_3} \propto \textrm{Re} [e^{-i\theta}e^{-i(2\pi B A_\textrm{area}/\Phi_0  -2\pi\bar{q}/\nu)}]$ and  $I^{\textrm{int}}_{\textrm{Ref, D}_3} \propto \textrm{Re} [e^{-i(2\pi B A_\textrm{area}/\Phi_0  -2\pi\bar{q}/\nu)}]$.
Note that the period of the magnetic-field dependence is determined by $\Phi_0 \equiv h/e$, rather than the anyon flux quantum $\Phi_0^*=h/e^*$, because the change of $N_L$ (the net number of anyonic quasiparticles minus anyonic quasiholes localized inside the interferometry loop) is allowed in the model of Supplementary Equation~\ref{NEWH}; see Ref.~\cite{Chamon,Kivelson}.
 In this case, the interference pattern of $I^{\textrm{int}}_{\textrm{D}_3}$ is shifted by $\theta$ from that of $I^{\textrm{int}}_{\textrm{Ref, D}_3}$, as either the magnetic field or the gate voltage varies.  
 
The Coulomb interaction parameters chosen for the left panel of Fig.~3a are $U_{\textrm{bulk}}=10 \hbar v_\textrm{p} / (2\nu L)$ and $U_{\textrm{int}}=0.1 \hbar v_\textrm{p} / (2\nu L)$, while $U_{\textrm{bulk}}=10 \hbar v_\textrm{p} / (2\nu L)$ and $U_{\textrm{int}}=0.9 \hbar v_\textrm{p} / (2\nu L)$ for the right panel.
 
Next, we consider a more general quantum Hall regime of filling factor $\nu' = \nu + \nu_0$ where $\nu = 1/ (2n+1)$ and $\nu_0$ is a nonzero integer, focusing on the situation that   
 the edge channels from the integer filling $\nu_0$ are fully transmitted through the QPCs while those from the fractional filling $\nu$ participate in the Fabry-Perot interference as in Fig 2. As shown below, the presence of the edge channels from the integer filling $\nu_0$ does not modify our main finding. Namely, in this case  the interference-pattern phase shift between $I^{\textrm{int}}_{\textrm{Ref, D}_3}$ and $I^{\textrm{int}}_{\textrm{D}_3}$ is identical to the phase $\theta$ of the $\nu_0 = 0$ case discussed in Equation~(6) of the main text. 
 
The Hamiltonian of this more general situation is the same as that in Supplementary Equation~\ref{NEWH} but with the replacement of $Q_{\textrm{bulk}} = \nu B A_{\textrm{area}}/\Phi_0 +\nu N_L -\bar{q}$ into $Q_{\textrm{bulk}} = \nu' B A_{\textrm{area}}/\Phi_0 +\nu N_L -\bar{q}$. 
The replacement $\nu \to \nu' = \nu + \nu_0$ counts the fact that the total amount of electron changes occupying the Landau levels of the interferometer bulk region is now $\nu' B A_{\textrm{area}}/\Phi_0$. In the same way as before (see Supplementary Equation~\ref{bulkInte}), we obtain the reference signal
\begin{align}
\langle I^{\textrm{int}}_{\textrm{Ref, D}_3} \rangle \propto & e^{\frac{-\pi^2\left(1-\frac{2\nu L}{\hbar v_\textrm{p}} U_{\textrm{int}}\right)^2}{\beta ( U_\textrm{bulk}-\frac{2\pi\nu L}{\hbar v_\textrm{p}} U_{\textrm{int}}^2)}}
\frac{\textrm{Re}\left[e^{i2\pi(\frac{\nu_0 BA_{\textrm{area}}}{\Phi_0} -\bar{q})} \Theta_3 (\frac{\nu'\pi B A_{\textrm{area}}} {\nu\Phi_0} -\frac{\pi\bar{q}}{\nu} +\frac{i\pi^2 \left(1-\frac{2\nu L}{\hbar v_\textrm{p}} U_{\textrm{int}}\right)}{\beta\nu\left( U_\textrm{bulk} -\frac{2\pi\nu L}{\hbar v_\textrm{p}} U_{\textrm{int}}^2\right)}, \exp (\frac{-\pi^2 \beta^{-1}\nu^{-2} }{(U_\textrm{bulk}-\frac{2\pi\nu L}{\hbar v_\textrm{p}} U_{\textrm{int}}^2)} ) ) 
\right]}{\Theta_3 ( \frac{\nu'\pi B A_{\textrm{area}}}{\nu\Phi_0} -\frac{\pi \bar{q}}{\nu}, \exp (\frac{-\pi^2 \beta^{-1}\nu^{-2}}{(U_\textrm{bulk} -\frac{2\pi\nu L}{\hbar v_\textrm{p}}U_{\textrm{int}})} ) )}. \label{Expression_fig3_ref_out}
\end{align}
We first discuss the magnetic-field dependence of the reference signal.
In the Coulomb dominated regime of $U_\textrm{int} \simeq \hbar v_\textrm{p}/(2\nu L)$,
the reference signal becomes $I^{\textrm{int}}_{\textrm{Ref, D}_3}\propto\textrm{Re}[e^{2\pi i(\nu_0 B A_{\textrm{area}}/\Phi_0-\bar{q})}]$, showing magnetic-field dependent oscillations with the period determined by $\nu_0$. This result coincides with an expression obtained with a capacitive interaction model in Ref.~\cite{Halperin}; cf. Eqs. (24) and (25) of Ref.~\cite{Halperin}. 
In the pure Aharonov-Bohm regime of $U_{\textrm{int}}\simeq 0$ and when $\beta U_\textrm{bulk} \ll 1$, the reference signal becomes $I^{\textrm{int}}_{\textrm{Ref, D}_3}\propto\textrm{Re}[e^{2\pi i(\nu_0 B A_{\textrm{area}}/\Phi_0-\bar{q})}(1+\frac{1}{2} \exp(\frac{\pi^2(2\nu-1)}{\beta\nu^2 U_{\textrm{bulk}}})e^{-2\pi i(\nu'B A_{\textrm{area}}/(\nu\Phi_0)-\bar{q}/\nu)})]$. This also agrees with Ref.~\cite{Halperin}; cf. the two dominant terms of Eq. (24) in Ref.~\cite{Halperin}. For the integer quantum Hall regime of $(\nu', \nu_0) = (2,1)$, this expression becomes $I^{\textrm{int}}_{\textrm{Ref, D}_3}\propto\textrm{Re}[e^{- 2\pi i B A_{\textrm{area}}/\Phi_0}]$, which agrees with the experimental data in Ref.~\cite{Ofek} for Aharonov-Bohm oscillations in the Fabry-Perot interferometry in the bulk filling factor $\nu'=2$ and with one fully transmitting channel ($\nu_0=1$).
Next, we discuss the gate-voltage dependence of the reference signal.
The gate-voltage dependent part of the reference signal becomes
$I^{\textrm{int}}_{\textrm{Ref, D}_3} \propto \textrm{Re} [e^{-i2\pi(\bar{\gamma}-\nu_0\bar{\alpha}) V_\textrm{G}}]$ in the Coulomb dominated regime of $U_\textrm{int} \simeq \hbar v_\textrm{p}/(2 \nu L)$, 
and $I^{\textrm{int}}_{\textrm{Ref, D}_3}\propto\textrm{Re}[e^{-2\pi i(\bar{\gamma}-\nu_0\bar{\alpha})V_\textrm{G}}
(1+\frac{1}{2} \exp(\frac{\pi^2(2\nu-1)}{\beta\nu^2 U_{\textrm{bulk}}})e^{-2\pi i(\nu'\bar{\alpha}V_\textrm{G} /\nu-\bar{\gamma}V_\textrm{G}/\nu)})]$ in the pure Aharonov-Bohm regime of $U_{\textrm{int}}\simeq 0$ and $\beta U_\textrm{bulk} \ll 1$.
The gate-voltage dependences are again in good agreement with Ref.~\cite{Halperin}. 
   
We also obtain the main interference signal $I^{\textrm{int}}_{\textrm{D}_3}$ for the general case of $\nu' = \nu + \nu_0$,
\begin{align}
\langle I^{\textrm{int}}_{\textrm{D}_3}\rangle & \propto \frac{\textrm{Re}\left[e^{i2\pi(\frac{\nu_0 BA_{\textrm{area}}}{\Phi_0} -\bar{q})}e^{-i\theta} \Theta_3 (\frac{\nu'\pi B A_{\textrm{area}}} {\nu\Phi_0} -\frac{\pi\bar{q}}{\nu} +\frac{i\pi^2 \left(1-\frac{2\nu L}{\hbar v_\textrm{p}} U_{\textrm{int}}\right)}{\beta\nu\left( U_\textrm{bulk} -\frac{2\pi\nu L}{\hbar v_\textrm{p}} U_{\textrm{int}}^2\right)}, \exp (\frac{-\pi^2 \beta^{-1}\nu^{-2} }{(U_\textrm{bulk}-\frac{2\pi\nu L}{\hbar v_\textrm{p}} U_{\textrm{int}}^2)} ) ) 
\right]}{\Theta_3 ( \frac{\nu'\pi B A_{\textrm{area}}}{\nu\Phi_0} -\frac{\pi \bar{q}}{\nu}, \exp (\frac{-\pi^2 \beta^{-1}\nu^{-2}}{(U_\textrm{bulk} -\frac{2\pi\nu L}{\hbar v_\textrm{p}}U_{\textrm{int}})} ) )}. \label{Expression_fig3_out}
\end{align}
Again as in the previous case of $\nu_0 = 0$, Supplementary Equation~\ref{Expression_fig3_out} is identical to Supplementary Equation~\ref{Expression_fig3_ref_out} except $e^{-i \theta}$, and the phase shift $\theta$ satisfies the expression in Eq.~(6) (see the main text) in the semiclassical regime of $e^* V \gg k_B T \gtrsim \hbar v_\textrm{p}/L$. Therefore, $\theta$ is detectable also in the general case of $\nu' = \nu + \nu_0$ from the phase shift between the reference signal and the main signal.

\vspace{.7cm}

\paragraph*{\bf Supplementary Note 6 : Discussion about Eqs.~(3) and (4).} We discuss Equations~(3) and (4) of the main text, focusing on the physical interpretation of Type II which involves topological vacuum bubbles.
We repeat the expression for Type II in Eq.~(3),
\begin{eqnarray}
I_{\textrm{D}_3}^{\textrm{II}} &\propto& \gamma_1^2 \gamma_2 \gamma_3 (e^* V)^{2\nu-1}(k_B T)^{2\nu-1}
\frac{2L + C L_T}{\hbar v_\textrm{p}} e^{-2L/L_T} \sin^2\pi\nu \cos(2\pi\Phi/\Phi_0^{*}+\pi\nu), \nonumber 
\end{eqnarray}
The factors $(e^*V)^{2 \nu}$ and $(k_BT)^{2 \nu}$ arise from the tunneling of an anyon biased by $e^* V$ and that of a thermally excited anyon, respectively. 
Other factors $(e^*V)^{-1}$, $(k_BT)^{-1}$, and $e^{-2L / L_T}$ describe the reduction of wave packet overlap, hence the reduction of interference amplitude. In particular the last one, $e^{-2L / L_T}$, is the well-known dephasing factor. The factor $2L + C L_T$ comes from the phase space for anyon braiding of Type II. 
In the factor of $\sin^2 \pi \nu$, one power, $\sin \pi \nu$, originates from the braiding phase $2 \pi \nu$ and the vacuum bubble effect (see Supplementary Figure~2 and the next paragraph), while 
the other $\sin \pi \nu$ originates from the exchange between two thermally excited anyons. 
The other expressions of Types I-1, I-2, and III in Eq.~(4) can be understood in a similar way.

We below explain how the braiding phase  $2 \pi \nu$ effectively occurs in Type II process.
\begin{itemize}
\item The two interfering paths of a Type II process are shown in Supplementary Figures~2a and b.
\item A gauge describing the exchange-statistics phase of anyons between different edges (Edges 2 and 3) is chosen for tunneling operators at QPC2 and QPC3, as shown in Equation~(12) of the main text. We choose the gauge in parallel to the extended edge scheme in Fig.~4 of the main text. 
\item When the particle-like ``virtual'' anyon thermally excited on Edge 2 at QPC2 jumps to Edge 3 (see Supplementary Figure~2a), it gains the exchange-statistics phase factor $\zeta_a = e^{-i \pi \nu}$ as its location is exchanged with the voltage-biased ``real'' anyon injected from Edge 1; see Supplementary Figure~2c. In the same gauge, the particle-like ``virtual'' anyon thermally excited on Edge 2 at QPC3 in Supplementary Figure~2b gains no exchange-statistics phase factor, when it jumps to Edge 3.
\item After the jump, the spatial ordering of anyons on Edge 2 is different between Supplementary Figures~2a and b. For the paths 2a and b to interfere, the anyons (the particle-like real anyon and the hole-like virtual anyon) for example in Supplementary Figure~2b must be exchanged as in Supplementary Figure~2d, for them to have the same spatial ordering as Supplementary Figure~2a. This results in another exchange-statistics phase factor of $\zeta_b = e^{i \pi \nu}$ shown in Supplementary Figure~2d. 
\item The two factors lead to the winding phase factor $\zeta_a^* \zeta_b = e^{i 2 \pi \nu}$ in the interference signal. 
\item The phase factors $\zeta_a$ and $\zeta_b$ are derived from the commutation rules in Eqs.~(11) and (12).
\end{itemize}

We further discuss Eqs.~(3) and (4).
The phase shift $e^* V L/\hbar v_\textrm{p}$ in
$I_{\textrm{D}_3}^{\textrm{I-1}} \propto \cos(2\pi\Phi/\Phi_0^*+e^* V L/\hbar v_\textrm{p})$ originates from the average of dynamical phase $2 k L$ over the energy window of $e^* V$, where $\hbar k$ is anyon momentum. $I_{\textrm{D}_3}^{\textrm{I-2}}$ is proportional to $\cos(2\pi\Phi/\Phi_0^*)$ without the shift $\propto e^*V$.
It is because Type I-2 is the interference between the anyon of energy $e^*V$ injected from S$_1$ and a thermally excited anyon so that the energy window for the average of the dynamical phase is determined by $\textrm{min} \{ e^* V, k_B T \} = k_B T$. $I_{\textrm{D}_3}^{\textrm{II}} \propto \cos(2\pi\Phi/\Phi_0^* + \pi \nu)$ and
$I_{\textrm{D}_3}^{\textrm{III}} \propto \cos(2\pi\Phi/\Phi_0^* - \pi \nu)$ do not have the phase shift $\propto e^*V$, because the interference occurs by thermally excited anyons.

We point out that the phase shift $\theta$ is well described by Eq.~(6) (see the main text) in the regime of $e^* V \gg k_B T, \hbar v_\textrm{p} / L$ (beyond $e^* V \gg k_B T \gg \hbar v_\textrm{p} / L$). It is because Types II and III share the common physics except only for the phase space ($2L + C L_T$ and $C L_T$, respectively) for anyon braiding. 
